\documentclass[aps,nofootinbib,superscriptaddress,twocolumn,10pt,prd]{revtex4-1}

\usepackage{graphicx}
\usepackage{amsmath,amssymb}
\usepackage{hyperref}
\usepackage{braket}
\usepackage{subfigure}
\usepackage{float}
\usepackage[dvipsnames]{xcolor}
\usepackage{soul}
\usepackage{rotating}
\usepackage{multirow}

\newcommand{\data}[1] {{\it #1}}

\hypersetup{
    colorlinks=true,
    linkcolor=red,
    citecolor=blue,
}

\newcommand{\joint}{{12}}

\allowdisplaybreaks

\begin{document}

\title{Concordance and Discordance in Cosmology}

\author{Marco Raveri}
\affiliation{Kavli Institute for Cosmological Physics, Department of Astronomy \& Astrophysics, Enrico Fermi Institute, The University of Chicago, Chicago, IL 60637, USA}
\author{Wayne Hu}
\affiliation{Kavli Institute for Cosmological Physics, Department of Astronomy \& Astrophysics, Enrico Fermi Institute, The University of Chicago, Chicago, IL 60637, USA}

\begin{abstract}
The success of present and future cosmological studies is tied to the ability to detect discrepancies in complex data sets within the framework of a cosmological model. Tensions caused by the presence of unknown systematic effects need to be isolated and corrected to increase the overall accuracy of parameter constraints, while discrepancies due to new physical phenomena need to be promptly identified.
We develop a full set of estimators of internal and mutual agreement and disagreement, whose strengths complement each other.
These allow to take into account the effect of prior information and compute the statistical significance of both tensions and confirmatory biases.
We apply them to a wide range of state of the art cosmological probes and show that these estimators can be easily used, regardless of model and data complexity.
We derive a series of results that show that discrepancies indeed arise within the standard $\Lambda$CDM  model.
Several of them exceed the probability threshold of $95\%$ and deserve a dedicated effort to understand their origin.
\end{abstract}

\maketitle

%
\section{Introduction}
%
Since the discovery of cosmic acceleration~\cite{Perlmutter:1998np,Riess:1998cb}, the description of our universe based on General Relativity with a cosmological constant ($\Lambda$) and cold dark matter (CDM) has provided  a successful working model for cosmology.
The success of the $\Lambda$CDM model relies on its ability to describe a wide array of different cosmological observations ranging from the spectrum of fluctuations in the Cosmic Microwave Background (CMB) to the clustering of galaxies and gravitational lensing observables.

Nevertheless discrepancies exist between the determination of $\Lambda$CDM parameters by different data sets.  
Local measurements of the Hubble constant differ from the value inferred from CMB observations of the Planck satellite~\cite{Ade:2015xua} by more than $3.4\,\sigma$~\cite{Riess:2016jrr}.
Measurements of the galaxy weak lensing correlation function also show disagreement with Planck CMB observations, involving parameters that determine the amplitude of the weak lensing signal, with a statistical significance that ranges between $1.7\, \sigma$ and $2.3\, \sigma$ for the Dark Energy Survey~\cite{Abbott:2017wau} and the Kilo Degree Survey~\cite{Hildebrandt:2016iqg} respectively.
Furthermore the internal consistency of the Planck CMB spectra in both temperature and polarization was analyzed in~\cite{Ade:2015xua,Aghanim:2016sns,Motloch:2018pjy} revealing some discrepancies between the temperature spectrum and the reconstruction of its lensing signal.

The existence of such discrepancies is in large part due to the advent of precision cosmology and the low statistical errors of large surveys.  
When facing these and other discrepancies, we have to understand whether they can be attributed to residual systematic effects, an incorrect modeling of the observables or new physical phenomena.
The next generation of cosmological probes, like Euclid~\cite{Laureijs:2011gra}, LSST~\cite{Abell:2009aa} and CMB-S4~\cite{Abazajian:2016yjj}, are expected to further raise experimental sensitivity.   
While these may resolve current controversies, their increased modeling complexity will also make it difficult to inspect the data sets or the parameter posteriors to identify future discrepancies. 
This will make it increasingly difficult to understand whether data sets agree or not and a failure at doing so will compromise their scientific return.

In this paper we discuss several concordance/discordance estimators (CDEs) that can be used to understand the internal consistency of a data set and its agreement with other cosmological probes.
First, inspired by the Bayesian evidence as a measure of goodness of fit, we introduce a test that exploits the statistics of the likelihood at maximum posterior.
Its dependence on the prior distribution allows to properly take into account data and prior constrained directions when counting degrees of freedom, while being significantly easier in practical applications with respect to the evidence.
Second, we study the statistics of the evidence ratio test of data set compatibility in order to understand its biases.
We show that in practical applications the bias toward agreement of the evidence ratio test is usually as large as its nominal value making its interpretation on the Jeffreys' scale unreliable in determining agreement or disagreement.
Third, we then define an estimator based on the ratio of likelihoods at maximum posterior, which maintains a close relationship with the evidence ratio in limiting cases, but allows for an easy assessment of statistical significance of the reported results.
Finally, we consider estimators that quantify shifts in the parameters of two data sets, providing an implementation that works in arbitrary number of dimensions and priors.

These tools can be straightforwardly applied, regardless of data and model complexity, and are based on a Gaussian linear model for the data likelihood and the posterior distribution that can be easily checked.
In addition they are  sensitive to both tensions between data sets and the presence of confirmatory biases.

We illustrate their application on current data sets and analyze known discrepancies between state of the art cosmological probes.
More specifically, we investigate the internal consistency of CMB measurements, establishing a set of benchmark results for the next release of the Planck data and showing that:
the cross correlation of the CMB temperature and E-mode polarization is a bad fit to the $\Lambda$CDM model due to the likely presence of residual, frequency dependent, systematics or foregrounds; the discrepancy between the CMB spectra and lensing reconstruction is present for both the temperature spectrum and the E-mode polarization spectrum, at about the same statistical significance; the measurement of the large angular scale CMB fluctuations is in tension with the small scale temperature and E-mode spectra with at a statistical significance of about the $95\%$ confidence level.

We recover the known tensions between CMB and local measurements of $H_0$ and weak lensing probes showing that the latter are slightly larger than what reported in literature, when considering the Canada-France-Hawaii Telescope Lensing Survey and the Kilo Degree Survey on large linear scales.
This tension is also slightly larger than what we estimate by looking at the $S_8\equiv \sigma_8\Omega_m^{0.5}$ parameter since this is not one of the principal components of both parameter covariances while our estimator optimally weights all parameter space directions.
We find that the CMB is in tension with probes of the clustering of galaxies, which can be attributed to the SDSS LRG DR4 survey being too good of an internal fit to different values of cosmological parameters.

This paper is organized as follows. 
In Sec.~\ref{Sec:Methodology} we discuss the technical aspects of several CDEs and their application to data.
In Sec.~\ref{Sec:Datasets} we detail the cosmological model and data sets and apply the CDEs to them in Sec.~\ref{Sec:Results}.
We summarize our conclusions in Sec.~\ref{Sec:Conclusions}.  
In a series of Appendices~\ref{App:QuadraticForms}-\ref{App:Tables}, we derive the statistical properties of the CDEs and give details on their implementation.

%
\section{Concordance discordance estimators} \label{Sec:Methodology}
%
In this section we introduce and review the Concordance/Discordance Estimators (CDEs) that we later apply to cosmological data sets.
We loosely refer to a CDE as a statement about a data set $D$ or a collection of data sets $D=D_1 \cup \dots \cup D_n$, within a given model $\mathcal{M}$, that  quantifies agreement or disagreement between the data and the model.
In case of a single data set these statements should quantify internal consistency (or self-consistency), in case of of multiple data sets, mutual consistency.

Since we regard data as random, CDEs are random variables as well, distributed over the space of data $D$.
When defining a CDE:
\begin{itemize}
\item we must be able to compute the distribution of the CDE over the space of data realizations $D$, where $D$ can be a single data set or the union of multiple data sets $D=D_1 \cup \dots \cup D_n$, depending on the definition of the CDE.
\item we  report the probability $P({\rm CDE}>{\rm CDE}_{\rm obs})$ so that low (high) probabilities identify disagreement (agreement) based on the observed value ${\rm CDE}_{\rm obs}$.
\end{itemize}
The distribution over data space is usually high dimensional and, though it is in principle possible to understand it with Monte Carlo techniques, doing so is typically extremely computationally intensive. For this reason we shall apply, and test the validity of, Gaussian approximations to work out analytically the distribution of these estimators.

Once probabilities over data space are computed, if $P({\rm CDE}>{\rm CDE}_{\rm obs})$ is too low then this could point toward the presence of tensions and if it is too high, the presence of confirmatory biases.  
Note that confirmation bias in this sense does not necessarily mean  a voluntary human action directed at confirming prior beliefs but includes any subtle assumption that can bias results toward accepting a fiducial model. 
These could include, as an example, overestimating data covariances, assuming a fiducial cosmology in the data reduction (e.g.~converting angles and redshifts to distances), calibrating  numerical algorithms around a given cosmology, and others.
As experimental precision increases, even subtle biases, if not properly counterbalanced would damage the scientific return of the affected experiment.

Many of the commonly employed estimators are presented in the literature without computing their statistics and rather interpreting their observed value as an indication of agreement/disagreement.
This does not take into account that CDEs can be biased, $\langle {\rm CDE} \rangle_D \neq 0$, toward agreement or disagreement.
Knowing the distribution over the data space prevents us from being tricked into thinking that there is agreement or disagreement when it is not the case. 

We next warn the reader about caveats in interpreting CDEs. 
CDEs can indicate agreement or disagreement but do not reveal the cause.  
In particular in case of tensions these could result from a problem with the data and unknown systematic effects, a problem with the predictions that stems from an incomplete modeling of the observable or a more fundamental problem with the model.
CDEs do not discriminate one from the other but  rather quantify the statistical level of unknowns in the given theoretical and experimental situation. 

Another limitation, common to all methods to quantify agreement/disagreement within a model is that they do not quantify the need for model extensions.     
It is always possible to relax tension with the addition of extra parameters, that could be describing systematic effects or new physical aspects of the model, but doing so carries the danger of over-fitting.
The methods that we describe in this paper should not be used to justify model extensions directly but rather motivate further studies with the appropriate statistical tools, like Bayesian model selection.
 
Just as no one CDE gives the probability of the model given the data, not all  CDEs  result in the same assessment of statistical significance for concordance
or discordance.
There are multiple ways in which the model can be in tension or agreement with the data.
In fact if the CDE is selected after looking at the data, one can always find some aspect of the data that deviates from the model just by chance fluctuations.
It is therefore advantageous to select, before looking at the data, multiple CDEs that correspond to meaningful quantities whose values we would want to be probable given a model.

Finally, when looking at these multiple CDE results, we should not naively combine them into a global probability.
To assess that, we would need to know the joint distribution of the multiple tests.
For example the CDEs might be correlated making multiple concordance or discordance results redundant.   
Even if the CDEs are uncorrelated we would expect that out of many tests, one might fail due to chance fluctuations.  
We instead use the  CDEs to flag individual aspects of the data and model for further study and multiple CDEs to assess the robustness of conclusions from any single CDE.

%
\subsection{Basic definitions}\label{Sec:BasicDefinitions}
%
We now lay out some definitions to clarify the notation of the subsequent sections. 

We  commonly employ the multivariate Gaussian distribution, over the space of $\theta$, that we denote as:
\begin{align} \label{Eq:GaussianDistribution}
\mathcal{N}_{N}(\theta ; \bar{\theta}, \mathcal{C}) = (2\pi)^{-N/2} |\mathcal{C}|^{-1/2} e^{ -\frac{1}{2} \left( \theta -\bar{\theta}\right)^T \mathcal{C}^{-1} \left( \theta -\bar{\theta}\right)} \,,
\end{align}
where $\det(\cdot) \equiv |\cdot|$, $N$ corresponds to the number of dimensions, $\bar{\theta}$ is the mean of the distribution and $\mathcal{C}$ is the covariance. Generally through this paper we  denote parameter covariances as $\mathcal{C}$ and data covariances as $\Sigma$. 
Given a model $\mathcal{M}$ and data $D$, the probability of the $N$ model parameters $\theta$ after the data $D$ is given by:
\begin{align} \label{Eq:Posterior}
P(\theta|D,\mathcal{M}) = \frac{ P(D|\theta,\mathcal{M}) P(\theta | \mathcal{M})}{P(D | \mathcal{M})} = \frac{\mathcal{L}(\theta) \Pi(\theta) }{\mathcal{E}} \,,
\end{align}
that we  call the parameter's posterior and where $P(\theta|\mathcal{M}) \equiv \Pi(\theta)$ is the prior probability density (pdf), normalized to unity over parameter space, $ P(D|\theta,\mathcal{M})\equiv\mathcal{L}(\theta)$ is the likelihood and $P(D|\mathcal{M})\equiv \mathcal{E}$ is the evidence. 
Usually the normalization of the posterior is not computed and one has to work with the following:
\begin{align} \label{Eq:UnNormalizedPosterior}
\mathcal{P}(\theta) \equiv \mathcal{L}(\theta) \Pi(\theta) \,,
\end{align}
that we  call un-normalized posterior. 
The normalization factor of the un-normalized posterior is the evidence:
\begin{align} \label{Eq:Evidence}
\mathcal{E} \equiv P(D|\mathcal{M}) = \int \mathcal{P}(\theta) \,d\theta = \int \mathcal{L}(\theta) \Pi(\theta) \,d\theta \,.
\end{align}
Notice that, within a given model $\mathcal{M}$, the evidence defines the prior probability for observing data $D$.
This is especially relevant in cosmology where we do not have the possibility of having truly different data realizations. 
Thus we have to fix the model that then predicts its distribution of data realizations that would be drawn from its evidence.
In this sense we can define functions of the data $D$ and, within a model $\mathcal{M}$, we can compute their distributions and, for example, their average over data realizations as:
\begin{align} \label{Eq:ExpectedValueDataRealizations}
\langle f(D) \rangle_D = \int f(D) P(D|\mathcal{M}) dD \,,
\end{align}
where the measure over data space is the evidence of the model.

As for the prior distribution, we use four different functional forms, depending on the application of interest:
\begin{itemize}
\item {\it Flat prior:} given by a ``tophat" function $\Pi(\theta)=1/ V_\Pi$ when all the $n$-th parameters components are included between $\theta_{\rm max}^{(n)}$ and $\theta_{\rm min}^{(n)}$.
The prior volume is $V_\Pi = \prod_{n=1}^{N} \left[ \theta_{\rm max}^{(n)}-\theta_{\rm min}^{(n)} \right]$. 
\item {\it Uninformative flat prior:} a flat prior where the range is chosen so that the prior is uninformative with respect to the data, i.e. $( \theta_{\rm max}^{(n)}-\theta_{\rm min}^{(n)})^2 \gg  \mathcal{C}_{\theta^{(n)}\theta^{(n)}}$.
\item {\it Gaussian prior:} given by a multivariate Gaussian $\Pi(\theta) = \mathcal{N}_N( \theta; \theta_\Pi, \mathcal{C}_\Pi )$ with mean $\theta_\Pi$ and covariance $\mathcal{C}_\Pi$. These priors are normalized to unity and their maximum value at $\theta_\Pi$ is $(2\pi)^{N/2} |\mathcal{C}_{\Pi}|^{1/2}$.
\item {\it Delta prior:} a Gaussian prior in the limit ${\cal C}_\Pi \rightarrow 0$ or 
$\Pi(\theta) = \delta(\theta -\theta_\Pi)$, a rather stubborn choice which we use for
pedagogical purposes.
\end{itemize}
Uninformative flat priors results can be related to  the Gaussian ones by appropriately setting the center parameter and in the limit $\mathcal{C}_{\Pi} \gg \mathcal{C}$. 
Moreover the Gaussian prior volume, that is formally undefined, can be taken as $V_\Pi = (2\pi)^{N/2}|\mathcal{C}_\Pi|^{1/2}$ to retain the same normalization as $\Pi(\theta) =1/V_\Pi$  at the peak.

Flat priors are the ones that are used in most practical applications but, to the best of our knowledge, it is not possible to derive simple analytic results in general. 
For this prior choice some directions in parameter space might be constrained by the data. When this is the case they  become uninformative flat priors where analytic results can be derived.
When they are much more informative than the data, on the other hand, their effect is closer to the delta prior case. 
For the intermediate, partially informative, case we approximate flat priors with Gaussian priors, taking into account that, when the prior needs to be directly evaluated, it would give $\Pi(\theta) =1/V_\Pi$.
This allows us to appreciate the two most important features of  flat priors: the shift between the maximum likelihood and the maximum likelihood as constrained by the prior; the information content of the prior, as modeled by the covariance of the bounded flat distribution $\mathcal{C} = ( \theta_{\rm max}^{(n)}-\theta_{\rm min}^{(n)})^2/12$.
For practical applications we discuss the Gaussian approximation of the MCMC sampled posterior in Appendix~\ref{App:GaussianApproximationMCMC}.

%
\subsection{Gaussian linear model} \label{Sec:GLM}
%
To understand the statistics of the CDEs discussed in this section we need to make some simplifying assumptions.
We  assume that the likelihood of the data is Gaussian distributed in data space and we  expand our model predictions to linear order in their parameter dependence. This results in the Gaussian linear model (GLM), that was discussed in~\cite{Seehars:2014ora,Seehars:2015qza}, and whose treatment we  mostly follow.
The assumptions of the GLM are somewhat restrictive but find many applications in cosmology. Most of the available data likelihoods are Gaussian distributions in the data and many probes, notably the CMB, constrain the parameters of the $\Lambda$CDM model sufficiently well that the linear approximation is
valid.
\begin{figure}[t]
\centering
\includegraphics[width=\columnwidth]{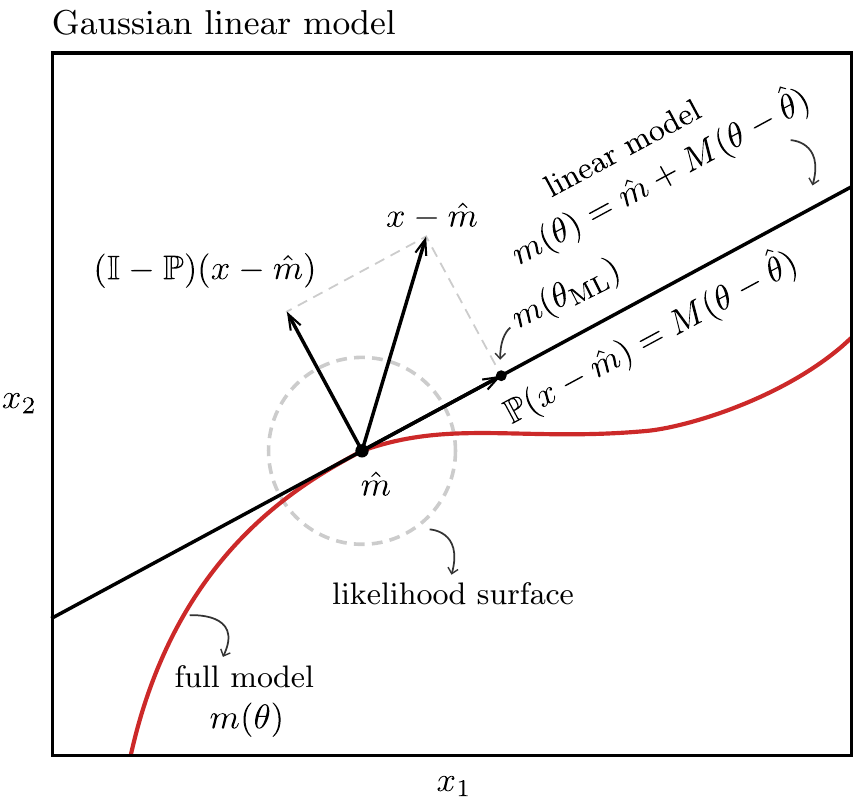}	
\caption{
Geometrical interpretation of the Gaussian linear model.
$(x_1,x_2)$ represents data space and $m(\theta)$ a one dimensional model, i.e.~a curve in the $(x_1,x_2)$ space. 
The figure also shows the linearization of the model and how to decompose differences between a data realization and the model (at fixed parameters) in the direction that is parallel and orthogonal to the model. $m(\theta_{\rm ML})$ shows the model corresponding to the best fit parameter values for the given data realization.
The dashed line shows a constant likelihood surface, where we assumed for simplicity that data covariance is proportional to the identity matrix.
}
\label{Fig:GaussianLinearModel}
\end{figure}
Let us assume that we have $d$ Gaussian distributed data points $x$, with mean $m$ and covariance $\Sigma$. Their likelihood is a Gaussian distribution in data space:
\begin{align} \label{Eq:GaussianDataLikelihood}
\mathcal{L} = \mathcal{N}_{d}(x ; m, \Sigma) \,.
\end{align}
Our model $\mathcal{M}$ would predict $m$ as a function of $N$ parameters $\theta$. We thus expand in series the prediction around a given parameter value $\hat{\theta}$:
\begin{align} \label{Eq:PredictionSeriesExpansion}
m(\theta) =\,& m(\hat{\theta}) +\left. \frac{\partial m}{\partial \theta}\right|_{\hat{\theta}} (\theta -\hat{\theta}) + \dots  \nonumber \\
\equiv\,& \hat{m}+M(\theta -\hat{\theta})+\ldots \,,
\end{align}
where we defined our central value for the expansion $\hat{\theta}$, the corresponding data prediction $\hat{m}=m(\hat{\theta})$ and the Jacobian of the transformation between data and parameter space $M$. 
The properties of the Jacobian are worth commenting. 
Since the dimension of the parameter space and data space is usually different, the Jacobian is not square and thus not invertible.
We can however define:
\begin{align} \label{Eq:GLMJacobianInverse}
\tilde{M}=(M^T\Sigma^{-1}M)^{-1}M^T\Sigma^{-1} \,,
\end{align}
that has the following properties:
\begin{itemize}
\item $\tilde{M}^T = \Sigma^{-1}M(M^T\Sigma^{-1}M)^{-1}$ given that $M^T\Sigma^{-1}M$ is symmetric;
\item $\tilde{M}M= M^T\tilde{M}^T = \mathbb{I}_{N \times N}$.
\end{itemize}
The two matrices $M$ and $\tilde{M}$ can be used to define a projector on the $m(\hat{\theta})$ tangent space:
\begin{align} \label{Eq:GLMProjector}
{\mathbb{P}} = M\tilde{M} \,,
\end{align}
with properties:
\begin{itemize}
\item ${\mathbb{P}}^2 = {\mathbb{P}}$, i.e. ${\mathbb{P}}$ is a projector and its complement is $\mathbb{I}-{\mathbb{P}}$;
\item ${\mathbb{P}} M\theta = M\theta$, leaves the tangent space of $m(\hat{\theta})$ invariant;
\item $(\mathbb{I}-{\mathbb{P}}^T)\Sigma^{-1} {\mathbb{P}}=0$ so that the complementary projectors are orthogonal in the metric defined by $\Sigma^{-1}$.
\end{itemize}
By decomposing the data residual $(x-m)$ in a component that is projected along the model, $\mathbb{P}(x-m)$, and a component that is orthogonal to the model,  $(\mathbb{I}-\mathbb{P})(x-m)$, we can now recast Eq.~(\ref{Eq:GaussianDataLikelihood}) into:
\begin{align} \label{Eq:GLMLikelihood}
\mathcal{L} = \mathcal{L}_{\rm max}\exp\left[ -\frac{1}{2}(\theta-\theta_{\rm ML})^T \mathcal{C}^{-1} (\theta-\theta_{\rm ML}) \right] \,, 
\end{align}
with maximum likelihood:
\begin{align} \label{Eq:GLMMaximumLikelihood}
\mathcal{L}_{\rm max} = \frac{\exp\left[ -\frac{1}{2}(x-\hat{m})^T( \mathbb{I} -{\mathbb{P}} )^T \Sigma^{-1} ( \mathbb{I} -{\mathbb{P}} ) (x-\hat{m})\right]}{(2\pi)^{d/2} |\Sigma|^{1/2} }   \,, 
\end{align}
maximum likelihood parameters and covariance:
\begin{align} \label{Eq:GLMLikelihoodParameters}
\theta_{\rm ML} =\, & \hat{\theta} +\tilde{M}(x-\hat{m}) \,, \nonumber \\
\mathcal{C}=\, & (M^T\Sigma^{-1}M)^{-1} \,.
\end{align}
Notice that the maximum likelihood parameter value depends on the data realization $x$.
Fig.~\ref{Fig:GaussianLinearModel} summarizes the geometrical meaning of the GLM in a two dimensional data space with a one parameter model.
\begin{figure*}[!ht]
\centering
\includegraphics[width=\textwidth]{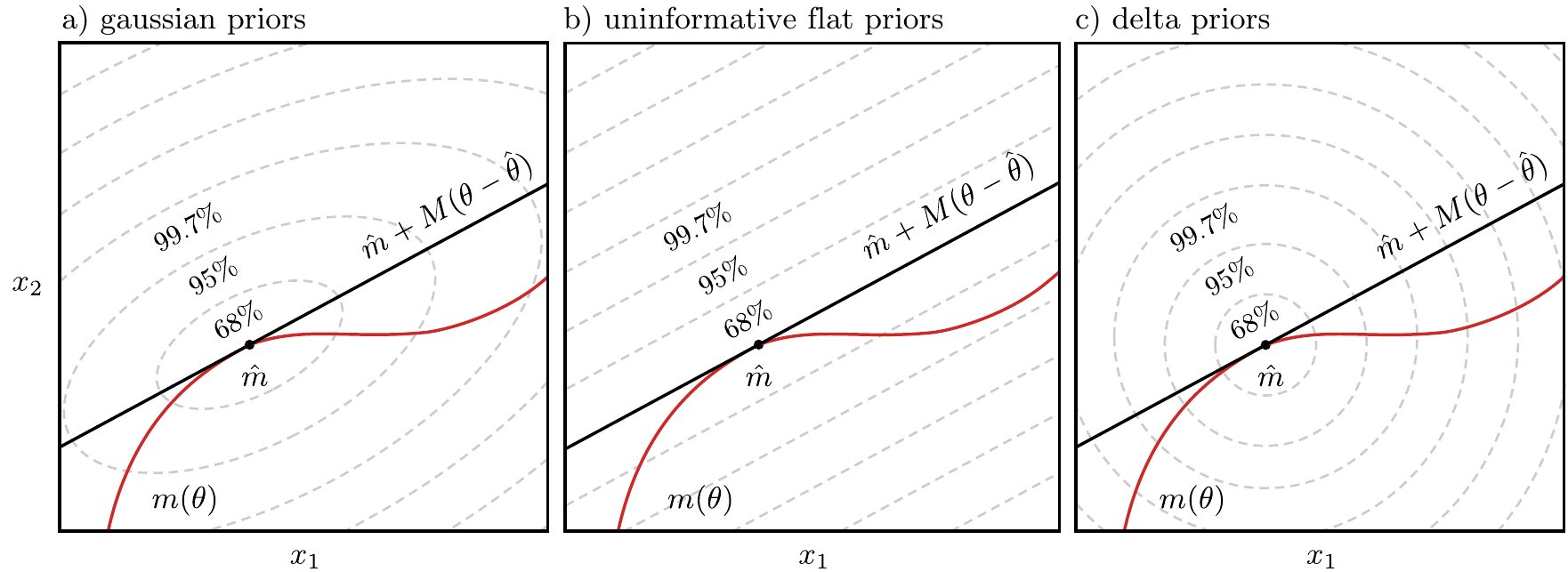}	
\caption{
Geometrical interpretation of the GLM evidence.
In all panels $(x_1,x_2)$ represents data space and $m(\theta)$ a one dimensional model, i.e. a curve in the $(x_1,x_2)$ space.
The figure also shows the linearization of the model. The dashed lines correspond to the evidence contours, for different prior choices, and different confidence levels.
The contours are showing that, when drawing data realizations from the evidence, these will be $68\%$ of the time inside the $68\%$ contour,  $95\%$ of the time inside the $95\%$ contour and so on.
As in the previous figure we assumed, for simplicity, that data covariance is proportional to the identity matrix. In the Gaussian prior case we also assumed that $m_{\Pi}=\hat{m}$.
}
\label{Fig:GaussianLinearModelEvidence}
\end{figure*}

Having computed the likelihood we can get the posterior of the data, for the GLM, with different prior choices.
In the case of Gaussian priors the posterior is still Gaussian $P(\theta|D,\mathcal{M}) = \mathcal{N}_N(\theta;\theta_p, \mathcal{C}_p)$ with 
\begin{align} \label{Eq:GLMGaussianPosterior}
\mathcal{C}_p =&\, (\mathcal{C}^{-1}_\Pi + \mathcal{C}^{-1} )^{-1} = (\mathcal{C}^{-1}_\Pi +M^T\Sigma^{-1}M)^{-1}  \,, \nonumber \\
\theta_p =&\, \mathcal{C}_p \left[ \mathcal{C}_\Pi^{-1}\theta_{\Pi} +\mathcal{C}^{-1}\theta_{\rm ML} \right] \nonumber \\
=&\, \mathcal{C}_p \left[ \mathcal{C}_\Pi^{-1}\theta_{\Pi} +M^T\Sigma^{-1}(x-\hat{m}+M\hat{\theta})\right] \,.
\end{align}
If we consider uninformative flat prior, 
 then the posterior is Gaussian $P(\theta|D,\mathcal{M}) = \mathcal{N}_{N}(\theta; \theta_{\rm ML}, \mathcal{C})$. In case of delta prior instead the posterior is a delta function around the chosen parameter value $P(\theta|D,\mathcal{M}) = \delta(\theta -\hat{\theta})$. 

The evidence can now be computed in a given model and for a given prior choice.
In parameter space and for Gaussian priors the evidence is given by:
\begin{align} \label{Eq:GLMGaussianEvidence}
\ln \mathcal{E} =&\, \ln \mathcal{L}_{\rm max} +\frac{1}{2}\ln \frac{|\mathcal{C}|}{|\mathcal{C}+\mathcal{C}_{\Pi}|} \nonumber \\
&\, -\frac{1}{2	}(\theta_{\rm ML} -\theta_{\Pi} )^T	( \mathcal{C}+\mathcal{C}_{\Pi} )^{-1}(\theta_{\rm ML} -\theta_{\Pi} ) \,,
\end{align}
where the first line contains the familiar Occam's razor term and the second line a penalty for cases where the prior center is not the maximum of the likelihood.
We can equivalently express this in terms of the likelihood
evaluated at the maximum posterior probability point $\theta=\theta_p$
\begin{align} \label{Eq:GLMGaussianEvidenceMaxPosterior}
\ln \mathcal{E} &= \ln {\mathcal L}(\theta_p) + \frac{1}{2} \ln | \mathcal{C}_p|  + \frac{N}{2}\ln (2\pi) + \ln\Pi(\theta_p)  \,.
\end{align} 
This form also highlights the limit which coincides with the case of uninformative flat priors where $\theta_{p}=\theta_{\rm ML}$, ${\cal C}_p={\cal C}$ and $\Pi(\theta_p)=1/V_\Pi$:
\begin{align}  \label{Eq:GLMFlatEvidence}
\ln \mathcal{E} =&\, \ln \mathcal{L}_{\rm max} +\frac{1}{2}\ln |\mathcal{C}| +\frac{N}{2}\ln (2\pi) -\ln V_\Pi  \,.
\end{align}
Likewise it highlights the delta prior limit, $\theta_p=\theta_\Pi$ where $\ln \mathcal{E} = \ln \mathcal{L}(\theta_\Pi)$, which is the limiting case of Gaussian priors as the prior covariance goes to zero.

We can now write these results in data space by means of the GLM.	
Fig.~\ref{Fig:GaussianLinearModelEvidence} shows the graphical interpretation of the GLM evidence, for different prior choices, in our two dimensional example.
In the Gaussian prior case, shown in panel a) of Fig.~\ref{Fig:GaussianLinearModelEvidence}, the evidence is a Gaussian distribution in data space $\mathcal{E} = P(D|\mathcal{M}) = \mathcal{N}_{d}( x; m_\Pi ,\Sigma_0 )$ with
\begin{align} \label{Eq:GLMGaussianPosteriorEvidence}
m_\Pi =\,& m(\theta_\Pi) \,,\nonumber\\
\Sigma_0 =\, & \Sigma +M\mathcal{C}_{\Pi} M^T \,.
\end{align}
In the uninformative flat prior case, the evidence is a Gaussian distribution orthogonal to the projector
\begin{align} \label{Eq:GLMFlatPosteriorEvidenceExpanded}
\mathcal{E} \propto e^{ -\frac{1}{2} (x-\hat{m})^T (\mathbb{I}-\mathbb{P})^T \Sigma^{-1} (\mathbb{I}-\mathbb{P})(x-\hat{m})} \,,
\end{align}
with a normalization factor such that the distribution integrates to unity over the data space. 
Notice that we have not defined this with the corresponding normal distribution since the
projection operation is not invertible so that the determinant of 
 $(\mathbb{I}-\mathbb{P})^T\Sigma^{-1}(\mathbb{I}-\mathbb{P})$ is singular.
If we consider delta priors, as in panel c) of Fig.~\ref{Fig:GaussianLinearModelEvidence}, the evidence is still Gaussian in data space $\mathcal{E} = \mathcal{N}_{d}(x;\hat{m},\Sigma)$. 

When studying the distribution of different quantities over data realizations, the evidence provides
the distribution of the data.
It is then a noteworthy result that, within a given model $\mathcal{M}$, regardless of the parameters, for all the considered prior choices, 
the data realizations provide an evidence that is a Gaussian distribution, with different mean and covariance. 

The last aspect of the GLM that we discuss is the dependence of the results on the expansion point $\hat{\theta}$.
When using the GLM to compute the distribution of different estimators the results do not depend on the arbitrary expansion point unless we purposely make that point special by prior choice.
 
%
\subsection{Goodness of fit type tests} \label{Sec:GoF}
%
The first application of the GLM consists in defining goodness of fit (GoF) type measures.
These are the only CDEs that we consider that measure the internal consistency of a single data set, within a model.

 We first define the usual maximum likelihood GoF measure as the  quadratic form:
\begin{align}\label{Eq:QML}
Q_{\rm ML} 
=\,& (x-\hat{m})^T( \mathbb{I} -{\mathbb{P}} )^T \Sigma^{-1} ( \mathbb{I} -{\mathbb{P}} ) (x-\hat{m}) \,,
\end{align}
Note that
\begin{align} \label{Eq:MaxLikelihoodGOF}
Q_{\rm ML}  =
-2 \ln\mathcal{L}_{\max} +2\langle \ln\mathcal{L}_{\max}\rangle_{D}+ \langle Q_{\rm ML} \rangle_D \,,
\end{align} 
where the average is over data realizations and so up to these constant offsets $Q_{\rm ML}$ is equivalent to $-2 \ln\mathcal{L}_{\max}$, the familiar effective $\chi^2$ at its minimum.
  This quadratic form therefore quantifies the distance between the data and the model at its best parameter point.
Taken as a CDE, if $P( Q_{\rm ML} > Q_{\rm ML}^{\rm obs})$ is too low then the data is a bad fit to the model and conversely if it is too high it is too good a fit to the data,  possibly indicating the presence of confirmatory biases. 

Eq.~(\ref{Eq:QML}) defines a quadratic form over data space and its distribution in general depends on the evidence, as the probability of data given the model,
which in turn depends on the prior.
However in this case the projection $ \mathbb{I} -{\mathbb{P}}$ in $Q_{\rm ML}$ 
makes its statistical properties independent of the prior and given by 
$Q_{\rm ML} \sim \chi^2(d-N_{\mathcal{ L}})$ (see App.~\ref{App:ProofsGoF}; here and below $\sim$ denotes distributed as).
Here  $N_{\mathcal{L}}={\rm rank} [\mathbb{P}]$ to take into account the fact that the likelihood might not be sensitive to some parameters if $\partial m(\theta) /  \partial \theta = 0$. If there are no irrelevant parameters $N_{\mathcal{L}}=N$.

Implicit in the use of $Q_{\rm ML}$ as a goodness of fit statistic is that the likelihood is maximized over all the relevant parameters without reference to or bounds from the prior. However, once the allowed model parameters are constrained by priors, we must adopt a different goodness of fit statistic.

The prior distribution usually encodes physical requirements on the model, like $\Omega_m \geq 0$, or a vague integration of previous experimental knowledge, like $20 \leq H_0\, [{\rm km}\,{\rm s}^{-1}{\rm Mpc}^{-1}] \leq 100$. 
We would not be interested in a model that fits well the data while violating physical requirements or accepted previous results.
The effect of the prior is to penalize such situations.

To define a GoF measure that takes the effect of the prior into account we start from the evidence. 
To see why, allow us to consider a one parameter ($\theta$) model and a data set that is directly measuring that parameter.
The evidence is then $\mathcal{E}=\int_{-\infty}^{+\infty} \mathcal{L}(D|\theta)\Pi(\theta)\,d\theta$. Under the simplifying assumption that the likelihood depends on the difference between the parameter and the data (that in this example is just the measured value of the parameter) the evidence, as a function of the data, becomes $\mathcal{E}=\int_{-\infty}^{+\infty} \mathcal{L}(\theta-D)\Pi(\theta)\,d\theta$.
This is the convolution integral that gives the probability density of the difference between the prior and the data.

The evidence GoF is then defined by analogy to Eq.~(\ref{Eq:MaxLikelihoodGOF}) as
\begin{align} \label{Eq:EvidenceGoFGeneral}
Q_{\mathcal{E}} & \equiv -2 \ln \mathcal{E} +2\langle \ln \mathcal{E} \rangle_{D} +  \langle  Q_{\mathcal{E}} \rangle_D \,.
\end{align}
Unlike $Q_{\rm ML}$, the specific quadratic form $Q_{\mathcal{E}}$ describing the data dependence of the evidence depends on the prior and so we give its explicit form for the various cases
below.
This statistics quantifies the compatibility between the prior and the likelihood, defining a goodness of fit statistics that is effectively conditioned on the prior.
We then apply the GLM to Eq.~(\ref{Eq:EvidenceGoFGeneral}) and for different prior choices. 

If we consider uninformative flat priors, the evidence quadratic form is given by:
\begin{align} \label{Eq:EvidenceGoFFlat}
Q_{\mathcal{E}} =&\, (x-\hat{m})^T(\mathbb{I}-\mathbb{P})^T\Sigma^{-1}(\mathbb{I}-\mathbb{P})(x-\hat{m}) \,,
\end{align}
just like $Q_{\rm ML}$ and is chi square distributed with $d-N_{\mathcal{L}}$ degrees of freedom.
This means that the evidence and maximum likelihood GoF statistics are identically distributed in case of uninformative flat priors as one might expect.

At the other extreme are delta priors. The evidence goodness of fit is determined by:
\begin{align} \label{Eq:EvidenceGoFDelta}
Q_{\mathcal{E}} = (x-m_\Pi )^T\Sigma^{-1}(x-m_\Pi) \,,
\end{align}
where $x \sim \mathcal{N}_{d} (x; m_\Pi, \Sigma)$ so that $Q_{\mathcal{E}} \sim \chi^2(d)$.
Notice that degrees of freedom counting is different than in the  uninformative flat prior case because the model cannot be optimized over the parameter space.
 
For Gaussian priors we have that:
\begin{align} \label{Eq:EvidenceGoFGaussian}
Q_{\mathcal{E}} =  (x-m_\Pi)^T(\Sigma +M\mathcal{C}_{\Pi} M^T)^{-1}(x-m_\Pi) \,.
\end{align}
Since the distribution of data draws is Gaussian, $x \sim \mathcal{N}_{d} (x; m_\Pi,\Sigma +M\mathcal{C}_{\Pi} M^T)$, 
$Q_{\mathcal{E}} \sim \chi^2(d)$ just like the delta prior case.
Although the model can now be optimized over the parameter space, $Q_{\mathcal{E}}$ pays a compensating penalty from the prior.

These results for the evidence highlight two aspects that are worth commenting.
The first is that the evidence GoF is the optimal estimator to weight differences between the prior and the data.
In both the delta and Gaussian prior cases the difference between the model with priors and the data draws $x-m_\Pi$ is weighted with its inverse covariance. We discuss in App.~\ref{App:OptimalQuadForm} what makes inverse covariance weighting optimal.
In case of uninformative flat priors, where there is no sense of preferred model parameters, this reduces to usual maximum likelihood GoF.
The second aspect is that there is a direct relationship between the evidence GoF and maximum likelihood based GoF that is the result of a hidden symmetry. 
We can always regard priors as external data so that the evidence GoF for Gaussian priors is the same as the maximum likelihood GoF if we add an additional data point for each Gaussian prior.
With Gaussian priors on all $N$ parameters, the maximum likelihood GoF would be distributed with $(d+N)-N=d$ degrees of freedom, as the evidence GoF.

In practical applications we want to define a GoF measure that retains the best properties of both the maximum likelihood GoF and the evidence GoF.
As the former measure we want it to be easy to compute while accounting for limitations that the prior places on optimizing parameters, that the latter measures.

Similar considerations in the literature for assessing Bayesian goodness of fit, for the purpose of model selection, has lead to the use of the deviance information criterion (DIC), which measures the improvement of the likelihood, within the region of support of the prior, relative to the number of effective parameters that the data constrain.
The DIC is defined as~\cite{RSSB:RSSB353,Kunz:2006mc,Liddle:2007fy,Trotta:2008qt,RSSB:RSSB12062}, 
\begin{align} \label{Eq:DIC}
{\rm DIC} \equiv -2 \ln \mathcal{L}({\theta}_p) +2N_{\rm eff} \,,
\end{align}
where ${\theta}_p$ is an estimate of the true parameters.
$N_{\rm eff}$ is the Bayesian complexity:
\begin{align}
N_{\rm eff} & \equiv  2 \ln \mathcal{L}(\theta_p) -2 \langle \ln \mathcal{L} \rangle_\theta  \,,
\end{align}
where the average is over the posterior. $\theta_p$ could be fixed to be the parameter means or the maximum point of the posterior.  
Note that with the commonly used flat priors, the maximum likelihood point within the prior range is the maximum of the posterior.
We therefore take the latter case for generality.

Now we can define by analogy to Eq.~(\ref{Eq:MaxLikelihoodGOF}) a new GoF statistic
\begin{align}
Q_{\rm MAP} = &\,  -2 \ln \mathcal{L}(\theta_p) 
+2\langle \ln \mathcal{L}(\theta_p) \rangle_{D} +  \langle  Q_{\rm MAP} \rangle_D ,
\end{align}
for the likelihood at the maximum a posteriori (MAP) point.
Since the specific quadratic form  for $Q_{\rm MAP}$ depends on the prior, we now consider
each case separately.

In both the delta and uninformative flat prior cases the likelihood at maximum posterior is distributed as the evidence  $Q_{\rm MAP} = Q_{\mathcal E}$.  
In the Gaussian prior case it defines a quadratic form in data space:
\begin{align} \label{Eq:MaxPostGauss}
Q_{\rm MAP} 
=&\, (x-m_\Pi)^T\bigg[ (\mathbb{I}-\mathbb{P})^T \Sigma^{-1} (\mathbb{I}-\mathbb{P}) \nonumber \\
&\, +\tilde{M}^T\mathcal{C}_{\Pi}^{-1} \mathcal{C}_p \mathcal{C}^{-1} \mathcal{C}_p \mathcal{C}_\Pi^{-1} \tilde{M} \bigg] (x-m_\Pi) \,.
\end{align}
This case also illuminates the meaning of $N_{\rm eff}$.
If some directions in parameter space are not constrained by the data, as it happens in many practical applications, the quadratic form defined by 	Eq.~(\ref{Eq:MaxPostGauss}) is lower rank, i.e.\ the model cannot invest all its nominal parameters in improving the goodness of the fit.
$Q_{\rm MAP}$ is distributed as a sum of Gamma distributed variables and its distribution can be conservatively approximated by that of a chi squared distributed variable with $d- {\rm tr}[ (\mathcal{C}_\Pi+\mathcal{C})^{-1}\mathcal{C}_{\Pi} ]$ degrees of freedom.
The trace term is $N_{\rm eff}$ under GLM  with Gaussian priors
\begin{align}
N_{\rm eff} 
&= N -{\rm tr}[ \mathcal{C}_\Pi^{-1}\mathcal{C}_p ].
\label{Eq:Neff}
\end{align} 
It can be interpreted as the effective parameters that a data set is constraining.
To see why, consider the limiting cases.  If the prior covariance is much wider than
the data covariance, this expression returns the full number of parameters $N$
whereas in the opposite limit where all parameters are prior limited it returns zero.  
Thus for any type of prior, $0 \le N_{\rm eff} \le N$ making the uninformative flat and
delta cases bounds on $N_{\rm eff}$ and limits of the statistics of $Q_{\rm MAP}$.

For the case of  flat priors which may be informative we can follow a similar procedure
of identifying the effective number of parameters using Eq.~(\ref{Eq:Neff}).   While
this approximation is not exact, it tends to be conservative.  Furthermore, 
being conservative for directions that are weakly constrained by the data mitigates
non-Gaussianity in the posterior.   Along these directions,  it is more likely that the posterior is non-Gaussian and with slowly decaying tails.

To summarize, our procedure gives the exact distribution  of $Q_{\rm MAP}$ for all parameter space directions that are either completely constrained by the prior or the data and in these limiting cases reduces to the evidence GoF. Moreover in case of completely data constrained parameters it further reduces to the maximum likelihood GoF measure.

%
\subsection{Evidence ratio type tests} \label{Sec:EvidenceRatio}
%
We next proceed to the application of the GLM to estimators that aim at quantifying the compatibility of data set couples.
One that has been applied in literature is the evidence ratio estimator of data set compatibility~\cite{Marshall:2004zd,Feroz:2008wr,March:2011rv,Amendola:2012wc,Verde:2013wza,Martin:2014lra,Karpenka:2014moa,Raveri:2015maa,Joudaki:2016mvz,Abbott:2017wau}.

With the posterior distribution of two different data sets we want to test whether they can be described with the same set of cosmological parameters.
This amounts to comparing the probabilities of two different statements:
\begin{itemize}
\item $\mathcal{I}_0$: the two data sets are described by the same choice of unknown parameters;
\item $\mathcal{I}_1$: the two data sets are described by independent choices of unknown parameters;
\end{itemize}
then we compute their probabilities and compare them:
\begin{align} \label{Eq:EvidenceRatio}
\mathrm{C} =\,& \frac{P(D_1 \cup D_2 | \mathcal{I}_0, \mathcal{M} )}{P(D_1 \cup D_2 | \mathcal{I}_1, \mathcal{M} )} \nonumber \\
=\,& \frac{P(D_1 \cup D_2 | \mathcal{M} )}{P(D_1 | \mathcal{M} )P(D_2 | \mathcal{M} )} \,,
\end{align}
where $P(D_1 \cup D_2 | \mathcal{M} )$ is the joint evidence of the two data sets while $P(D_1 | \mathcal{M} )$ and $P(D_2 | \mathcal{M} )$ is the evidence for the single ones.
Since we are working with two data sets, $D_1$ and $D_2$, we  use the subscript $1$, $2$ and $\joint$ to indicate quantities referring to the first, the second and the joint data sets respectively.

Used in the form of Eq.~(\ref{Eq:EvidenceRatio}) the evidence ratio does not provide an estimate of the statistical significance of the reported results. It is common in the literature to  interpret the outcome on a Jeffreys' scale~\cite{Jeffreys61,kassr95}: $\ln \mathrm{C} < 0$ indicates tension between the data sets
and $\ln \mathrm{C} > 0$ agreement;  $3:1$ odds one way or the other is ``substantial'', $10:1$ is ``strong'', $30:1$ is ``very strong'', $100:1$ is ``decisive''.
This has the disadvantage that the Jeffreys' scale is not calibrated on the specific application at hand and using it might give misleading results~\cite{Amendola:2012wc,Nesseris:2012cq,Grandis:2016fwl}. 

In case of uninformative flat priors the Gaussian approximation for the evidence ratio can be immediately read from Eqs.~(\ref{Eq:GLMGaussianEvidence},\ref{Eq:GLMFlatEvidence}):
\begin{align} \label{Eq:EvidenceRatioWFPGLM}
\ln\mathrm{C} =\, & \ln\mathcal{L}_{\rm max}^{\joint} -\ln\mathcal{L}_{\rm max}^{1}-\ln\mathcal{L}_{\rm max}^{2} \nonumber \\
& +\frac{1}{2}\ln \frac{|\mathcal{C}_{\joint}|}{|\mathcal{C}_{1}||\mathcal{C}_{2}|} +\ln \frac{V(\Pi_{1})V(\Pi_{2})}{V(\Pi_{\joint})}   \nonumber \\
& -\frac{N_1+N_2 -N_{\joint}}{2}\ln (2\pi) \,,
\end{align}
and this shows that, when averaging this quantity over $D_1 \cup D_2$ realizations, several terms would not cancel out, i.e.~this CDE is biased.
In the case of uninformative flat priors the calculation explicitly gives:
\begin{align} \label{Eq:EvidenceRatioAverageWFPGLM}
\langle \ln\mathrm{C} \rangle_{\joint}	 =\, &  -\frac{N_1+N_2 -N_{\joint}}{2}[1+\ln (2\pi)] \nonumber \\
& +\frac{1}{2}\ln \frac{|\mathcal{C}_{\joint}|}{|\mathcal{C}_{1}||\mathcal{C}_{2}|} +\ln \frac{V(\Pi_{1})V(\Pi_{2})}{V(\Pi_{\joint})} \,.
\end{align}
Notice that, in practical applications, the Occam's razor factors in the second line of Eq.~(\ref{Eq:EvidenceRatioAverageWFPGLM}), are much larger than the first line, thus making the evidence ratio biased toward agreement since $\mathcal{I}_1$ effectively involves two Occam's factors compared with one for $\mathcal{I}_0$.

We define the debiased evidence ratio test as:
\begin{align} \label{Eq:DebiasEvidenceRatio}
\Delta \ln\mathrm{C} = -2\ln\mathrm{C}   +2\langle \ln\mathrm{C} \rangle_{\joint} \,.
\end{align}
If $\Delta \ln\mathrm{C}$ is significantly greater than zero, this indicates tension, if it is smaller than zero it indicates confirmation bias. 
The confidence level of the statement can be computed using the GLM.
The proofs of the results of this section can be found in App.~\ref{App:ProofsEvR}.

In case of uninformative flat priors, $\Delta \ln\mathrm{C}$ is, up to an additive constant, chi squared distributed with $N_1+N_2 -N_{\joint}$ degrees of freedom and the observed value can be read from Eqs.~(\ref{Eq:EvidenceRatioWFPGLM},\ref{Eq:EvidenceRatioAverageWFPGLM}).
In case of delta priors the evidence ratio is trivially distributed as $\Delta \ln\mathrm{C}=0$ for all data draws.

For Gaussian priors, the distribution is more complicated and is, in general, a sum of independent variance-gamma distributed variables, see App.~\ref{App:ProofsEvR}.
Notice that in this case, to obtain the distribution of the Gaussian prior evidence ratio from that of the maximum likelihood ratio, treating the prior as additional data, we need to take into account the fact that we add the prior to the analysis of both $D_1$, $D_2$ and $D_{\joint}$.
If we now regard the prior as data, since the prior is not changing in the analysis of the different data sets, data draws of $D_1$ and $D_2$ would be correlated by the prior. The evidence in the Gaussian prior case is then in correspondence with the maximum likelihood ratio of correlated data sets.

As with GoF in the previous section, we aim at defining a CDE that retains ease of use as the ratio of maximum likelihoods, that does not require heavy use of numerical integration to compute the statistical significance, like the evidence ratio, but at the same time encodes the effect of the prior.
This suggests that we again examine the statistics of the various likelihoods  at their maximum posterior point. 

We therefore consider the difference of log-likelihoods at their MAP point
\begin{equation} \label{Eq:LikeRatioMP}
Q_{\rm DMAP} \equiv  -2\ln {\mathcal L}_{\joint}(\theta_p^{\joint}) +2\ln {\mathcal L}_1(\theta_p^1) +2\ln {\mathcal L}_2(\theta_p^2)  \,.
\end{equation}
Note that in this case the normalization factors in ${\mathcal L}$, which provide the offset mean values in
Eq.~(\ref{Eq:MaxLikelihoodGOF}), drop out of the difference so long as $1$ and $2$ are independent data sets.
If data are drawn from the evidence with uninformative flat priors and delta priors the distribution of $Q_{\rm DMAP}$ is the same as the distribution of the evidence ratio. In the Gaussian prior case its distribution  is conservatively approximated with a chi squared distribution:
\begin{align} \label{Eq:LikeRatioMPGauss}
Q_{\rm DMAP} \sim \chi^2( N_{\rm eff}^{1}+N_{\rm eff}^{2}-N_{\rm eff}^{\joint}) \,.
\end{align}
Its exact distribution in terms of a sum of Gamma distributed variables, can be found in App.~\ref{App:ProofsEvR}.

This estimator quantifies the loss in goodness of fit when combining two data sets. When considering single data sets, the model parameters can be separately optimized within the prior; when joining them, there is less freedom in model parameter optimization. 
The ratio of likelihoods at maximum posterior tell us whether this decrease in goodness of fit is consistent with expectation from statistical fluctuations or not. 

The statistics of $Q_{\rm DMAP}$ is the same as the evidence ratio, once Occam's factors are removed, for completely data or prior constrained directions, while it differs over partially constrained directions. Over these the statistical significance of agreement/disagreement is underestimated as a mitigation strategy against non-Gaussianities.

This discussion allows us to shed light on the deviance information criterion (DIC) ratio estimator, as introduced in~\cite{Joudaki:2016mvz} to assess the agreement between CFHTLenS and Planck.   Using Eq.~(\ref{Eq:DIC}), we can define the DIC ratio
\begin{equation} \label{Eq:DICratio}
\ln \mathcal{I} = -\frac{1}{2} \bigg[ {\rm DIC}(D_1 \cup D_2) -{\rm DIC}(D_1)-{\rm DIC}(D_2) \bigg] \,.
\end{equation}
Similarly to the evidence ratio, Eq.~(\ref{Eq:DICratio}) is expected to indicate agreement or disagreement between two posterior distributions if it is found negative or positive respectively.
Depending on the evaluation point $\theta_p$ for DIC, the statistics of the DIC ratio changes accordingly.
If $\ln {\mathcal L}(\theta_p)$ in the DIC statistic is evaluated at the maximum posterior then twice the DIC ratio is distributed as $Q_{\rm DMAP}$, up to a data independent constant.
If the maximum likelihood is taken without regard for the prior, the distribution is chi squared, with $N_1+N_2-N_{\joint}$ degrees of freedom, similarly to the maximum likelihood ratio and the evidence ratio in the uninformative flat prior case.
This clarifies the relationship between the evidence ratio, the DIC ratio and $Q_{\rm DMAP}$.
When the data is either informative or completely uninformative these three quantities measure the same aspect of agreement/disagreement with different mean values over data space.

%
\subsection{Parameter differences} \label{Sec:ParametersQuad}
%
The next application of the GLM is to understand the distribution of quadratic forms in model parameters.
These are  natural generalizations of the usual rule of thumb estimator for tension and contain, as sub-cases, other estimators that have been proposed in literature.

If we consider two independent random variables $\theta_1$ and $\theta_2$ the probability density of their difference, in one dimension, $\Delta\theta \equiv \theta_1-\theta_2$, is given by the convolution integral of the two probability densities, $P_{\theta_1}$ and $P_{\theta_2}$, as:
\begin{align} \label{Eq:Exact1DShift}
P(\Delta\theta) = \int_{-\infty}^{+\infty} P_{\theta_1}(\tilde{\theta}) P_{\theta_2}(\tilde{\theta}-\Delta\theta)\,d\tilde{\theta} \,.
\end{align}
Tension between the measurements would be indicated if $P(\Delta\theta)$ has most of
its support at very negative or positive $\Delta\theta$.   
For the former case, there would be a low probability for the difference to be greater than zero:
\begin{align}
P( \Delta\theta>0 ) =&\, \int_{0}^{\infty}  P(\Delta\theta) \, d\Delta\theta \,.
\end{align}
To account for the possibility that the observed tension could be in either direction, we
take the smaller of $P( \Delta\theta>0 )$ and $P( \Delta\theta<0 )$.
The probability of obtaining a 1D parameter shift, $T_1$, more extreme than the data, in either direction, is then
\begin{equation}
\label{Eq:T1twotailed}
P(T_1>T_1^{\rm obs}) = 2\,{\rm min}\left[ P( \Delta\theta>0 ), P( \Delta\theta<0 ) \right] \,.
\end{equation}
We refer to this as the 1D parameter shift tension statistic. This holds for any two independent probability distributions and can be easily evaluated numerically.

If we assume that the two distributions, $P_{\theta_1}$ and $P_{\theta_2}$, are Gaussian then
we can evaluate this probability analytically.   Since the convolution of two Gaussians is another
Gaussian, with a variance given by the sums of the individual Gaussians, the
tension statistic becomes the usual ``rule of thumb difference in mean''. This consists in comparing the difference in the best fit values, or means, of one parameter for two different data sets to the quadrature sum of the parameters' variances: 
\begin{align} \label{Eq:1DDifferenceMean}
T_1(\theta) \equiv \frac{|\theta(D_1)-\theta(D_2)|}{\sqrt{\sigma_\theta^2(D_1)+\sigma_\theta^2(D_2)}} \,,
\end{align}
where $\theta(D_i)$ is the parameter best fit (or mean), for a given model and data set $D_i$, $\sigma^2_\theta(D_i)$ denotes its variance.
The statistical significance of the 1D parameter shift then becomes $P(T_1>T_1^{\rm obs})= {\rm Erf}(T_1/\sqrt{2})$, where ${\rm Erf}$ is the error function.

Because of its simplicity, this estimator is an easy and intuitive proxy to understand tensions between data sets and is also accurate if differences in a parameter are manifest at the posterior level.
However, there is no guarantee that the overall consistency of two generic data sets is properly signaled: the method needs to pick up the ``right'' parameter where all tension is expressed; it would not work right away in the multidimensional case; it does not take into account the effect of priors. 

When considering more than one dimension we can turn to the GLM to understand the statistics of tension estimators that, like the ``rule of thumb difference in mean'', are defined in parameter space.

We  consider differences in the posterior means of two different data sets:
\begin{align} \label{Eq:MeanDifference}
\Delta \bar{\theta} \equiv \theta_{p1}-\theta_{p2} \,,
\end{align}
that can be easily computed, from the results of Sec.~\ref{Sec:GLM}, as:
\begin{align} \label{Eq:DifferenceMeanGaussian}
\Delta \bar{\theta}  = \mathcal{C}_{p1}\left[ \mathcal{C}_\Pi^{-1} \theta_\Pi +\mathcal{C}_1^{-1}\theta^{\rm ML}_{1} \right]
-\mathcal{C}_{p2}\left[ \mathcal{C}_\Pi^{-1} \theta_\Pi +\mathcal{C}_2^{-1}\theta^{\rm ML}_{2} \right] \,, \nonumber \\
\end{align}
for Gaussian priors, and:
\begin{align}  \label{Eq:DifferenceMeanWFP}
\Delta \bar{\theta} = \theta^{\rm ML}_1-\theta^{\rm ML}_2 \,,
\end{align}
in case of uninformative flat priors. Note that under the GLM, the parameter means are the same as the parameters at the maximum 
posterior point.

Notice that both Eqs.~(\ref{Eq:DifferenceMeanGaussian}) and~(\ref{Eq:DifferenceMeanWFP}) are defined in terms of the parameters that $D_1$ and $D_2$ have in common, so that, when there are additional parameters describing systematic effects in one data set the corresponding distributions has to be marginalized over them.
When treating Gaussian priors, we  assume that the prior center is the same for both data sets and equal to the prior center of their combination, as we assumed in the previous sections.

Since the posterior means depend on the data, we now turn to the computation of the statistics of their differences over the space of joint data draws from $D_1$ and $D_2$.

Since $\Delta \bar{\theta}$ is a linear combination of correlated Gaussian variables, it is Gaussian distributed.
Furthermore, it can be shown that, for both Gaussian and uninformative flat priors:
\begin{align} \label{Eq:DeltaMean}
\langle \Delta \bar{\theta} \rangle_{\joint} = 0 \,.
\end{align}
Notice that this holds if the prior center is fixed (to an arbitrary value) for $D_1$, $D_2$ and $D_{\joint}$. If this is not the case and the prior center is different for the different data sets, the expectation value of the parameter difference is non-zero.

We are then left with computing the covariance of $\Delta \bar{\theta}$.
In case of uninformative flat priors this reads:
\begin{align}  \label{Eq:MLDeltaCovarianceWFP}
\mathcal{C}( \Delta \bar{\theta} ) = \mathcal{C}_{p1} + \mathcal{C}_{p2} \,,
\end{align}
while in case of Gaussian priors, 
 direct computation from the GLM gives:
\begin{align} \label{Eq:MeanDeltaCovarianceGauss}
\mathcal{C} ({\Delta \bar{\theta}}) = \mathcal{C}_{p1} + \mathcal{C}_{p2} - \mathcal{C}_{p1} \mathcal{C}_\Pi^{-1} \mathcal{C}_{p2} - \mathcal{C}_{p2} \mathcal{C}_\Pi^{-1} \mathcal{C}_{p1} \,.
\end{align}
These results can be directly obtained by means of the covariance of the joint data draws reported in App.~\ref{App:ProofsEvR}.

Having computed the distribution of $\Delta \bar{\theta}$, we can compute the distribution of a related quantity that carries the same information but has useful properties when applied in practice to data sets with non-Gaussian posteriors.
This is the difference between the mean parameters of one data set and the mean parameters of the joint data set.

We  refer to this quantity as the update difference in mean since it quantifies the differences in parameters of one data set when updating it with another one.
If we assume that the GLM applies to $D_1$ and $D_2$ then it also applies to $D_{\joint}$ and we can write the update difference in mean as:
\begin{align}
\Delta\bar{\theta}_{U} \equiv \theta_{p1}-\theta_{p\joint} = \mathcal{C}_{p1} (\mathcal{C}_{p1} + \mathcal{C}_{2})^{-1} (\theta_{p1} -\theta_{2} ) \,,
\end{align}
that still has zero mean and covariance:
\begin{align} \label{Eq:CovarianceUpdateML}
\mathcal{C} (\Delta\bar{\theta}_{U}) = \mathcal{C}_{p1} -\mathcal{C}_{p\joint} \,,
\end{align}
for both uninformative flat priors and Gaussian priors.
Since the CDEs discussed in this section are defined in terms of the parameter space posterior only, it is simple to derive all of the above results on parameter differences by considering the two data sets and the prior as independently measuring $\theta$ directly in parameter space using the projected covariances $\mathcal{C}$.

Notice that the previously discussed covariances have to be positive definite or positive semi-definite.
While this is true when all distributions are well defined Gaussian in the application to real data, with covariances from MCMC sampling, it might not be strictly true. We shall come back to the problem of computing this estimator in Sec.~\ref{Sec:ResParameters}. 

We are now in a position to define CDEs based on quadratic forms of parameter differences.
Given a positive semi-definite matrix $A$ we  define two types of quadratic estimators, depending on the vector that we use to define them.

If we consider $\Delta \bar{\theta}$ we have difference in mean quadratic CDEs defined as:
\begin{align} \label{Eq:DMQuadraticForm}
Q_{\rm DM} = (\Delta \bar{\theta} )^T A\, (\Delta \bar{\theta}  )  \,, 
\end{align}
while if we use $\Delta \bar{\theta}_U$ we have update difference in mean quadratic CDEs defined as:
\begin{align} \label{Eq:UDMQuadraticForm}
Q_{\rm UDM} = (\Delta\bar{\theta}_{U}  )^T A\,(\Delta\bar{\theta}_{U}  )  \,.
\end{align}
All these quadratic forms are central and some times degenerate, depending on the rank of $A$.

Belonging to this family of CDE we have two estimators that have been previously studied.
The first one is the difference in mean $\Delta \bar \theta$ with  $A=\mathcal{C}^{-1}(\Delta\bar{\theta})$ (e.g.~\cite{Battye:2014qga,Charnock:2017vcd,Lin:2017ikq}).
The second one is the surprise, that was introduced in~\cite{Seehars:2014ora} and used in~\cite{Seehars:2015qza,Grandis:2015qaa,Grandis:2016fwl}, and corresponds to  $\Delta \bar{\theta}$ with $A=\mathcal{C}_1^{-1}$ which is related to the Gaussian approximation of the Kullback-Leibler divergence~\cite{Kullback:1951va} between different data sets' posteriors.  The consideration of quadratic forms for the update $\Delta \bar{\theta}_U$ is new to this work as far as we are aware. 

For either $\Delta \bar \theta$ or $\Delta\bar{\theta}_{U}$, the optimal choice of $A$ is the inverse covariance  of the parameter difference that is being considered.
Other measures, provided that their distribution is properly calculated, can only underestimate rare events by not weighting them properly compared to an optimal measure.
We discuss the criterion that makes inverse covariance weighting optimal in App.~\ref{App:OptimalQuadForm}.   This also clarifies why the ``rule of thumb difference in mean'' in one dimension works so well when all the tension is manifest in one parameter where the choice in weighting of multiple dimensions is absent.

We therefore consider only $A= \mathcal{C}^{-1}$ in the following.   
With this choice the quadratic forms of Eqs.~(\ref{Eq:DMQuadraticForm},\ref{Eq:UDMQuadraticForm}) are chi squared distributed with degrees of freedom $\langle Q_{\rm DM} \rangle = {\rm rank}[ \mathcal{C}(\Delta\bar{\theta})]$ and $\langle Q_{\rm UDM} \rangle = {\rm rank}[ \mathcal{C}(\Delta\bar{\theta}_{U})]$ respectively.

%
\section{Model and data sets} \label{Sec:Datasets}
%
Our baseline model is the six parameter $\Lambda$CDM model as defined by:
cold dark matter density $\Omega_{c}h^2$; baryon density $\Omega_{b}h^2$; the angular size of the sound horizon $\theta_{\rm MC}$; the spectral index of the primordial spectrum of scalar fluctuations $n_s$ and its amplitude $\ln (10^{10}A_s)$; the reionization optical depth $\tau$. 
We also include in the model massive neutrinos, fixing the sum of their masses to the minimal value allowed by flavor oscillation measurements $\sum_{\nu} m_{\nu} = 0.06 \,\,\mbox{eV}$~\cite{Long:2017dru}.
We discuss in Appendix~\ref{App:Priors} the priors that we use throughout this work.

We analyze the level of agreement of several, publicly available, cosmological data sets within the $\Lambda$CDM model. 
The first data set that we consider consists of the measurements of CMB fluctuations in both temperature (T) and polarization (EB) of the {\it Planck} satellite~\cite{Ade:2015xua, Aghanim:2015xee}.
We further consider the {\it Planck} 2015 full-sky lensing potential power spectrum~\cite{Ade:2015zua} in the multipoles range $40\leq \ell \leq 400$. We exclude multipoles above $\ell=400$ as CMB lensing, at smaller angular scales, is strongly influenced by the non-linear evolution of dark matter perturbations. 

We include the ``Joint Light-curve Analysis'' (JLA) Supernovae sample~\cite{Betoule:2014frx}, which combines SNLS, SDSS and HST supernovae with several low redshift ones.
We also use BAO measurements of: BOSS in its DR12 data release~\cite{Alam:2016hwk}; the SDSS Main Galaxy Sample~\cite{Ross:2014qpa}; and the 6dFGS survey~\cite{Beutler:2011hx}.
We include the galaxy clustering power spectrum data derived from the SDSS LRG survey DR4~\cite{Tegmark:2006az} and the WiggleZ Dark Energy Survey galaxy power spectrum as measured from $170,352$ blue emission line galaxies over a volume of $1\,\mbox{Gpc}^3$~\cite{Drinkwater:2009sd,Parkinson:2012vd}.
For both data sets we exclude all the data points with $k>0.08\, h/{\rm Mpc}$.

We consider the measurements of the galaxy weak lensing shear correlation function as provided by the Canada-France-Hawaii Telescope Lensing Survey (CFHTLenS)~\cite{Heymans:2013fya} in their reanalyzed version of~\cite{Joudaki:2016mvz} and the Kilo Degree Survey (KiDS)~\cite{Hildebrandt:2016iqg}.
We applied ultra-conservative cuts, that make both CFHTLenS and KiDS data insensitive to the modelling of non-linear evolution and we included uncertainties in the modelling of intrinsic galaxy alignments, as in~\cite{Joudaki:2016mvz,Hildebrandt:2016iqg}. 
A posteriori we notice that, when considering ultra-conservative data cuts intrinsic alignment parameters are weakly constrained.

We use local measurements of the Hubble constant derived by the ``Supernovae, H0, for the Equation of State of dark energy'' (SH0ES) team~\cite{Riess:2016jrr} with the calibration of~\cite{2018arXiv180101120R}.
In addition we employ measurements derived from the joint analysis of three multiply-imaged quasar systems with measured gravitational time delays, from the H0LiCOW collaboration~\cite{Bonvin:2016crt}.

We combine the previously discussed data sets into families probing similar physical processes: a CMB family composed by CMB temperature, polarization and CMB lensing reconstruction; a ``background'' family joining supernovae and BAO measurements; the combination of SDSS LRG and WiggleZ measurements probing the clustering of galaxies; CFHTLenS and KIDS joined together in a Weak Lensing probe; Hubble constant's measurements from SH0ES and H0LiCOW.

Notice that the ``background'' family is not measuring the Hubble constant as SN measurements are analytically marginalized over intrinsic luminosity. 
The galaxy clustering data set is not measuring the present day amplitude of CDM perturbations $\sigma_8$ as both power spectrum measurements are separately marginalized over the power spectrum amplitude.
The H0LiCOW data set in turn does not only measure $H_0$ but a combination of $H_0$ and $\Omega_m$ since we implemented the full non-Gaussian likelihood described in~\cite{Bonvin:2016crt}.

Table~\ref{Table:Datasets} summarizes the data sets, acronyms and literature references for all the data sets used in this work.
We use the CAMB code~\cite{Lewis:1999bs} to compute the predictions for all the cosmological observables described above and we Markov Chain Monte Carlo (MCMC) sample the posterior of the previously discussed experiments with CosmoMC~\cite{Lewis:2002ah}.

Our results rest on two assumptions: linear theory modeling of the observables 
and the accuracy of the GLM.
As a sanity check of the former, we compare the parameter posterior and best fit prediction of the data, as obtained by neglecting and including non-linear modeling of the matter distribution, described by Halofit~\cite{Smith:2002dz}, with the updated fitting formulas described in~\cite{Takahashi:2012em,Mead:2015yca}.
We find that, with the above discussed set-up, the parameter posterior and best fits are not noticeably different.

\begin{table}
\setlength{\tabcolsep}{4pt}
\centering
\label{Table:Datasets}
\begin{tabular}{@{}lllc@{}}
\toprule
Acronym  & Data set & Year & Reference \\
\colrule
\data{lowl} & Planck low-$\ell$ TEB  & 2015 & \cite{Aghanim:2015xee} \\
\data{CMBTT} & Planck high-$\ell$ TT    & 2015 & \cite{Aghanim:2015xee} \\
\data{CMBEE} & Planck high-$\ell$ EE    & 2015 & \cite{Aghanim:2015xee} \\
\data{CMBTE} & Planck high-$\ell$ TE    & 2015 & \cite{Aghanim:2015xee} \\
\data{CMBL}   & Planck CMB Lensing & 2015 & \cite{Ade:2015zua} \\
\data{SN}       & JLA & 2014 & \cite{Betoule:2014frx} \\
\begin{tabular}{@{}l@{}} \data{BAO}  \\ $\,$ \end{tabular} & \begin{tabular}{@{}l@{}} BOSS DR12  \\ + SDSS MGS + 6dFGS \end{tabular}   & 
\begin{tabular}{@{}l@{}} 2011-15  \\ $\,$ \end{tabular} &
\begin{tabular}{@{}l@{}} \cite{Gil-Marin:2015nqa,Ross:2014qpa,Beutler:2011hx}   \\ $\,$ \end{tabular}  \\
\data{LRG}      & SDSS LRG survey DR4 & 2006 & \cite{Tegmark:2006az}  \\
\data{WiggleZ} & WiggleZ survey & 2012 & \cite{Drinkwater:2009sd,Parkinson:2012vd} \\
\data{CFHTLenS} & CFHTLenS survey & 2016 & \cite{Joudaki:2016mvz}    \\
\data{KiDS}         & KiDS survey         & 2016 & \cite{Hildebrandt:2016iqg} \\
\data{H}    & SH0ES                 & 2016 & \cite{Riess:2016jrr,2018arXiv180101120R}    \\
\data{HSL} 	 & H0LiCOW          & 2016 & \cite{Bonvin:2016crt} \\
\colrule
\begin{tabular}{@{}l@{}} \data{CMB} \\ $\,$ \end{tabular} & \begin{tabular}{@{}l@{}} \data{lowl} + \data{CMBTTTEEE}  \\ + \data{CMBL} \end{tabular}   & 
\begin{tabular}{@{}l@{}} 2015 \\ $\,$ \end{tabular} & \begin{tabular}{@{}l@{}} - \\ $\,$ \end{tabular} \\
\data{BG}          & \data{SN} + \data{BAO} & 2011-15 & - \\
\data{GC}        & \data{LRG} + \data{WiggleZ} & 2006-12 & - \\
\data{WL}         & \data{CFHTLenS} + \data{KIDS} & 2016 & - \\
\data{H0}           & \data{H} + \data{HSL} & 2016 & - \\
\botrule
\end{tabular}
\caption{Summary of data sets and data sets combinations used in this work.}
\end{table}

All the techniques considered in this work rely on the applicability of the Gaussian approximation to either the likelihood or the posterior of the considered data set.
Most of the considered data sets have Gaussian likelihoods, with the exception of the \data{HSL} and \data{lowl} data sets that we  exclude from tests requiring Gaussianity of the data likelihood.
We build the Gaussian approximation of the parameter space posterior and we check whether we can reliably use it, as discussed in Appendix~\ref{App:GaussianApproximationMCMC}.
We find that the posterior of all combinations of data sets containing the CMB power spectrum can be well approximated by Gaussian distributions in the parameters. Single weakly constraining data sets, on the other hand, usually result in non-Gaussian parameter posteriors.

%
\section{Application and results} \label{Sec:Results}
%
In this section we discuss the application of the CDEs in Sec.~\ref{Sec:Methodology} to cosmological data. 

Specifically to assess the internal consistency of a data set we consider the likelihood at maximum posterior as a goodness of fit measure:
\begin{align}
\label{Eq:MAPsummary}
Q_{\rm MAP}  &\, \equiv  -2 \ln \mathcal{L}(\theta_p) -d\ln(2\pi) -\ln(|\Sigma|) \nonumber\\
&\,  \sim \chi^2(d-N_{\rm eff}) \,, \nonumber \\
N_{\rm eff} &\, \equiv N -{\rm tr}(\mathcal{C}_{\Pi}^{-1}\mathcal{C}_p) \,.
\end{align}

To test the compatibility of data sets couples, $D_1$ and $D_2$, we consider the ratios of likelihoods at their maximum posterior:
\begin{align}
\label{Eq:DMAPsummary}
Q_{\rm DMAP} \equiv \,& -2\ln {\mathcal L}_{\joint}(\theta_p^{\joint}) +2\ln {\mathcal L}_1(\theta_p^1) +2\ln {\mathcal L}_2(\theta_p^2)  \nonumber \\
\sim\, & \chi^2( N_{\rm eff}^{1}+N_{\rm eff}^{2}-N_{\rm eff}^{\joint}) \,,
\end{align}
that measures the decrease in, prior constrained, goodness of fit when combining two data sets.

This is paired with parameter shifts in their update form:
\begin{align}
\label{Eq:UDMsummary}
Q_{\rm UDM} \equiv \,& (\theta_{p}^1 -\theta_{p}^{\joint})^T (\mathcal{C}_{p 1}-\mathcal{C}_{p \joint })^{-1} (\theta_{p}^{1} -\theta_{p}^{\joint}) \nonumber \\
\sim \,& \chi^2( {\rm rank}[\mathcal{C}_{p1 }-\mathcal{C}_{p\joint }] ) \,.
\end{align}
When possible we apply these  CDEs to every data set alone and to sets that define  families of physical probes, to test their internal consistency.
Then we  move to testing the consistency of different families by probing all their possible combinations.

Different tests applied to the same data sets provide complementary information that is helpful in singling out possible problems.
Goodness of fit type tests inform us of the internal consistency of the data sets but do not specifically highlight confirmation biases or tensions that look like parameters changes.
The ratios of likelihoods at their maximum posterior and parameter shifts tests on the other hand are designed to isolate problems along parameter modes.
In particular the former estimator is sensitive to shifts in all the parameters that two data set jointly constrain  while the latter is sensitive to shifts in the constraints that one of the data set improves over the other.

As an example, the goodness of fit test for a data set might fail, indicating a tension. Still parameter deviations, probed by the other two tests, might not be statistically significant, indicating that possible systematic effects or new physics is not mimicking the effect of a change in parameters.
In a cosmological context the matter power spectrum could be indicating the presence of an additional physical scale resulting in a scale dependent growth.
With sufficient experimental accuracy this will fail goodness of fit tests, as growth in the $\Lambda$CDM model is scale independent.
On the other hand this may not fail parameter shift tests as none of the nominal $\Lambda$CDM parameters can exactly describe this effect.
Conversely, a smooth dark energy component will generally result in a scale independent modification to the growth of structures that might mimic the effect of a change in $A_s$ or other cosmological parameters.
This will not show up at goodness of fit level but might show up at parameter level when we compare two probes that are differently sensitive to the amplitude of perturbations, for example measuring it at different redshifts.
In this case also the joint goodness of fit test is not guaranteed to fail as it might be dominated by the data set with larger number of data points.

In addition to these aspects, different tests, when applied in practice, have different responses to the presence of non-Gaussianities in the data and parameter spaces and thus have different failure modes. Testing multiple ones ensures that these are easily identified.
In particular if the posterior of a given experiment is non-Gaussian because the low probability tails decay slower than a Gaussian distribution the evidence ratio and parameter shift estimator have, different, opposite responses.
While the first one would overestimate tensions and underestimate confirmation the second one is built to mitigate this and may underestimate tensions.

In Appendix~\ref{App:Tables} we report, in table format, the full results of the application of the CDEs that we discuss to data.
In addition, we also report the results that can be obtained with the 1D parameter shifts and ``rule of thumb difference in mean'' statistics, when evaluated with our data configuration and analysis pipeline, to recover some known results that we  use as a benchmark for our estimators.

\subsection{Goodness of fit type tests} \label{Sec:ResGoF}
\begin{figure*}[!ht]
\centering
\includegraphics[width=\textwidth]{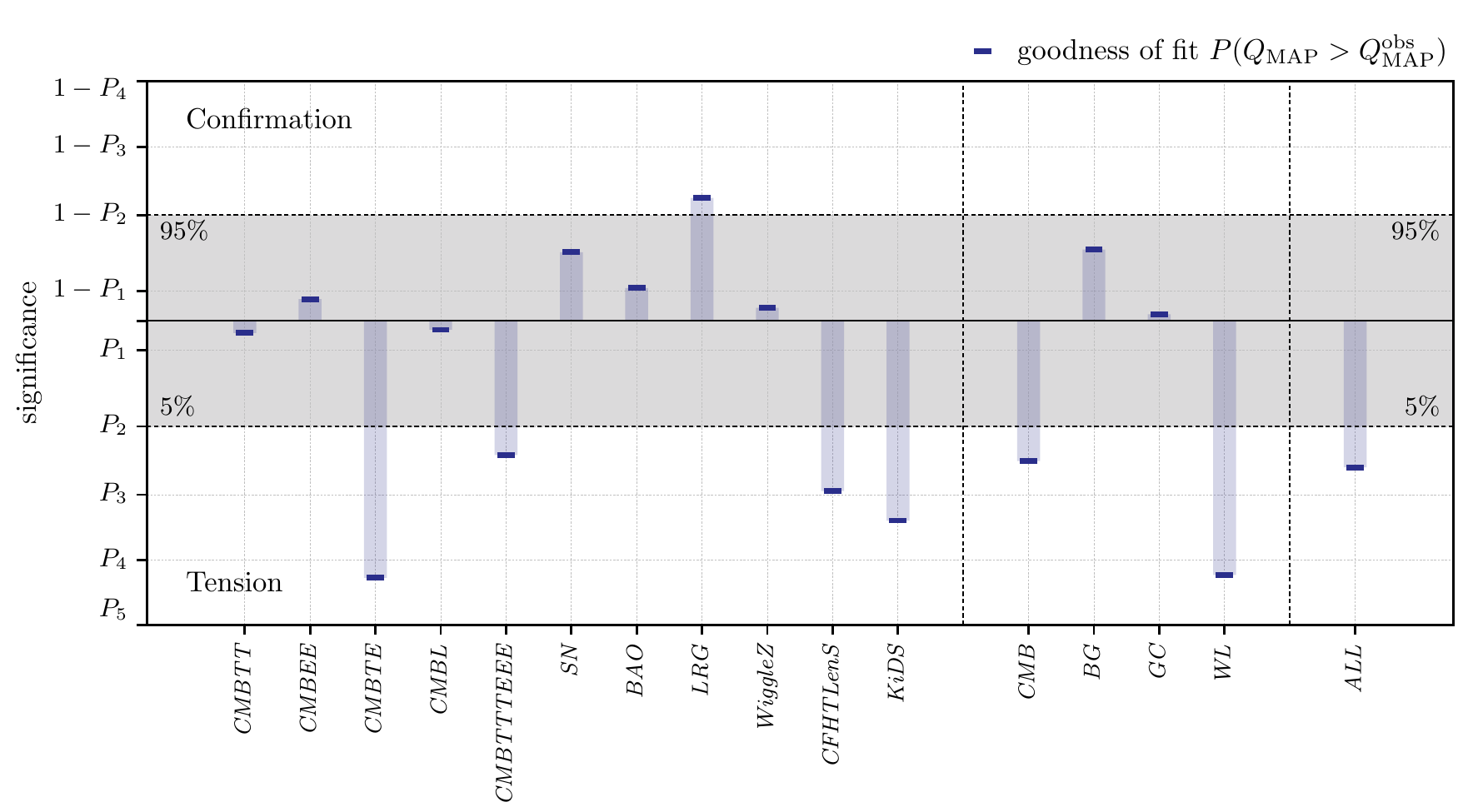}	
\caption{ \label{Fig:GoodnessOfFit}
The statistical significance of the posterior goodness of fit estimator, $Q_{\rm MAP}$ from Eq.~(\ref{Eq:MAPsummary}), applied to different data sets and data sets combinations.
The labels report different levels of statistical significance: $P_1\equiv 32\%$, $P_2\equiv 5\%$, $P_3\equiv 0.3\%$, $P_4\equiv 0.007\%$ and $P_5\equiv 0.00006\%$.
The darker shade indicates results that are not statistically significant.
}
\end{figure*}

In this section we present the application of the goodness of fit measures that were discussed in Sec.~\ref{Sec:GoF}.

In applying these estimators to real data there are two major challenges.
The first one consists in obtaining accurate best fit estimates. This involves global optimization of the posterior and is complicated by the large number of parameter space dimensions usually involved in cosmological studies.
What proves particularly challenging in this respect is the presence of mostly unconstrained parameters that can create multiple local maxima in the posterior.
This can be mitigated by having well converged MCMC parameter chains whose sample best fit estimate provides a good starting point to eventually find the global minimum with appropriate algorithms.

The second challenging aspect is to estimate correctly the number of parameters that a data set is constraining, $N_{\rm eff}$.
Prior distributions are in practice often non-Gaussian, for example when some direct or derived parameter 
is limited to be in a certain range.  Nonetheless in all cases, we adopt Eq.~(\ref{Eq:Neff})
for its calculation.   This is a reasonable approximation in that the comparison of the prior covariance
$\mathcal{C}_\Pi$ to the posterior covariance $\mathcal{C}_p$
always provides a criteria for when the prior is informative and the parameter cannot be optimized to the data.
As a concrete example, consider the tophat prior on a single parameter where $\mathcal{C}_{\Pi} = (\theta_{\rm max} -\theta_{\rm min})^2/12$.   Eq.~(\ref{Eq:Neff})
tells us that $N_{\rm eff}=1/2$ when the prior variance equals the data variance.
For the tophat prior, this occurs when the half-width is $\sqrt{3}$ times the rms of the data constraint, i.e.~between $1\sigma$ and $2\sigma$ of a Gaussian data constraint.

Therefore Eq.~(\ref{Eq:Neff}) suffices for an estimate even for this highly non-Gaussian prior so long as we allow for errors in each partially constrained direction 
at the level of a few tenths of a parameter. 
We have verified this error estimate with numerical simulations in one dimension, noticing that the error depends on the value of $N_{\rm eff}$:
it is small in the two limits $N_{\rm eff}=0$ and $N_{\rm eff}=1$ where the distribution is exact; 
increases as $N_{\rm eff}$ decreases from $N_{\rm eff}=0.9$ to $N_{\rm eff}=0.1$ approximately ranging from $0.1$ to $0.4$;
in this same range of $N_{\rm eff}$ the distribution of the $Q_{\rm MAP}$ estimator is increasingly conservative.

Evaluating Eq.~(\ref{Eq:Neff}) for  $N_{\rm eff}$ also requires well sampled parameter distributions to limit errors in parameter covariance estimates.
We thus require the Gelman and Rubin $R$ test~\cite{Gelman:1992zz,An98stephenbrooks} to satisfy $R-1<0.005$ for the worst constrained covariance eigenvalue.
We can then check sampling errors on the number of effective parameters as their variance across different MCMC chains and we find that these are usually of the same order as $R-1$.

In order to have a reliable estimate of $N_{\rm eff}$ we also need a good knowledge of the prior covariance.
This is built by joining different blocks. 
We directly MCMC sample the prior on the base $\Lambda$CDM parameters because of priors on derived parameters.
Flat priors on nuisance parameters are uncorrelated with priors on the base parameters and their diagonal entry in the prior covariance is built out of the covariance of the flat distribution. 
Some nuisance parameters have Gaussian priors that are uncorrelated with other priors. Their covariance entry can be easily set with the variance of the Gaussian prior.
Further details about the modeling of the prior distribution can be found in Appendix~\ref{App:Priors}.

Once these technical aspects have been properly addressed we can check the estimate of the number of effective parameters that a data set is constraining against physical intuition.
We list in Table~\ref{Table:NeffParams} the values of $N_{\rm eff}$ and the number of nominal parameters for the data sets that we consider.

\begin{table}[!ht]
\setlength{\tabcolsep}{6pt}
\centering
\begin{tabular}{@{}llllllll@{}}
\toprule
Data set & $N_{\rm eff}$ & $N$ \\
\colrule
\data{CMBTT} & 14.3 & 21 \\
\colrule
\data{CMBEE} &  8.1 & 13 \\
\colrule
\data{CMBTE} &  7.9 & 15 \\
\colrule
\data{CMBL} & 2.5 & 7 \\
\colrule
\data{SN} & 3.0 & 8 \\
\colrule
\data{BAO} &  3.1 & 6 \\
\colrule
\data{LRG} &  2.5 & 6 \\
\colrule
\data{WiggleZ} &  1.9 & 6 \\
\colrule
\data{CFHTLenS} & 1.8 & 7 \\
\colrule
\data{KiDS} & 1.8 & 7 \\
\botrule
\end{tabular}
\caption{ \label{Table:NeffParams}
The number of effective parameters, $N_{\rm eff}$, and the number of nominal parameters, $N$, for the different data sets that we consider.
}
\end{table}

As we can see the primary CMB spectra have seven, five and seven  parameters for \data{CMBTT}, \data{CMBEE} and \data{CMBTE} respectively
that are not constrained by the data.
These are nuisance parameters describing foregrounds and are instead constrained by informative Gaussian priors~\cite{Aghanim:2015xee}.  
CMB lensing has four unconstrained parameters $\tau$, $n_s$, $\Omega_b h^2$ and a calibration parameter. A combination of the other cosmological parameters, mainly  $A_s$ and $\Omega_c h^2$, is well constrained by the lensing amplitude whereas the directions constraining the shape of
the potential are only partially constrained.
\data{SN} constrain three parameters, the total matter  density $\Omega_m$ and two nuisance parameters, the intrinsic supernovae color and stretch.
The \data{BAO} data set constrains three parameters as it includes redshift space distortions measurements, so that only $\tau$ and $n_s$ are unconstrained while $A_s$ is mostly unconstrained.
The \data{LRG} and \data{WiggleZ} data sets constrain slightly more than two parameters, that are combinations of $\Omega_m$, $\Omega_b$ and $H_0$, thanks to the detection of the BAO feature in the matter power spectrum.  
Both \data{CFHTLenS} and \data{KiDS} constrain two parameters, the amplitude of the weak lensing signal and the amplitude of intrinsic alignment.
The latter, while not being detected, is slightly constrained over the prior and thus enters in degree of freedom counting.

The number of effective parameters that combinations of these data sets constrain is consistent with what we would expect from these results.
Notice that no physical knowledge was input to get the results of Table~\ref{Table:NeffParams} that automatically and accurately recover the physical results to a fraction of a parameter.

We can now turn to the probabilities associated with the values of $Q_{\rm MAP} $ in the various cases, as displayed in Fig.~\ref{Fig:GoodnessOfFit}.
In applying these estimators to the data we cannot use the \data{lowl} and \data{HSL} data sets as their likelihood is not Gaussian in the data points.
We have to exclude the \data{H} data set as the full data likelihood is not provided and we just have the parameter likelihood.

As we can see from both Fig.~\ref{Fig:GoodnessOfFit} and Table~\ref{Table:GoFresults} the \data{CMBTT}, \data{CMBEE},  \data{CMBL}, \data{SN}, \data{BAO}, \data{WiggleZ} data set are a reasonable fit to the data showing no tension nor confirmation at high statistical significance.
The \data{CMBL} result showcases the use of the maximum posterior as a goodness of fit measure. 
This data set has no irrelevant parameters and if we were to count all its parameters as being optimized this would indicate the presence, at a 5\% probability to exceed, of tensions. Since the $\Lambda$CDM model cannot use all its nominal parameters due to the priors, it is actually still a good fit to the \data{CMBL} data.

The \data{CMBTE} data set in turn is not a good fit at high statistical significance. The result is stable against degree of freedom counting since the goodness of fit, in this case, is dominated by the number of data points in the fit.
Since, as noted in~\cite{Aghanim:2015xee}, the coadded frequency spectrum is a good fit we suspect that this result is dominated by frequency dependent rather than cosmological effects, e.g.~foreground and systematics modeling, especially in the $100\, {\rm GHz }\times 217\, {\rm GHz }$ and $100\, {\rm GHz }\times 100\, {\rm GHz }$ spectra that have been highlighted in~\cite{Aghanim:2015xee}, at about the same statistical significance.

The full \data{CMB} goodness of fit is dominated by the TE results, whose statistical significance gets diluted by the increased number of data points in the joint data set. The results for the \data{CMBTTTEEE} data set further confirms this showing that the discrepancy in the fit cannot be attributed to \data{CMBL} measurements.
Moreover, the goodness of fit results for all data sets joined together (\data{ALL}) is dominated too by \data{CMB} results since this is the data set with the largest number of data points.

At slightly lower statistical significance we find that the \data{CFHTLenS} and \data{KiDS} data sets are a bad fit and the goodness of fit of their union further confirms this at high statistical significance.
Notice that this result is particularly worrisome since both data sets are cut at linear cosmological scales and thus should not be influenced by the, possibly improper, modeling of non-linearities.
The statistical significance of the goodness of fit to the joint \data{WL} data set is only slightly lower than the product of the single data sets, showing that the bad fits are almost independent.
These results could be, at least in the case of the \data{KiDS} data set, due to lack of modeling of survey geometry in the covariance, as reported in~\cite{Troxel:2018qll}. The same explanation does not apply to \data{CFHTLenS} whose covariance was obtained through simulations.

At a statistical significance that is borderline between significant and not significant we find that the \data{LRG} data set is confirmation biased.
Notice that, in this case, proper degree of freedom counting is crucial to the assessment of such effects.
If we were to assume that this data set measures all $\Lambda$CDM parameters this result will not be statistically significant.
If we further assume that the two bias parameters that have been marginalized over are also constrained by the data, the statistical significance of confirmation bias would decrease becoming $96\%$ for $N_{\rm eff}=3.5$ and $93\%$ for $N_{\rm eff}=4.5$.

Finally we notice that the \data{BG} data set is a good fit, while being dominated by the \data{SN} data set that has more data points with respect to the \data{BAO} one.
The same effect is seen for the \data{GC} data set where the statistical significance of confirmation in \data{LRG} measurements is over-weighted by the number of data points in the \data{WiggleZ} data set.

\subsection{Evidence ratio type tests} \label{Sec:ResEvidence}
\begin{figure*}[!ht]
\centering
\includegraphics[width=\textwidth]{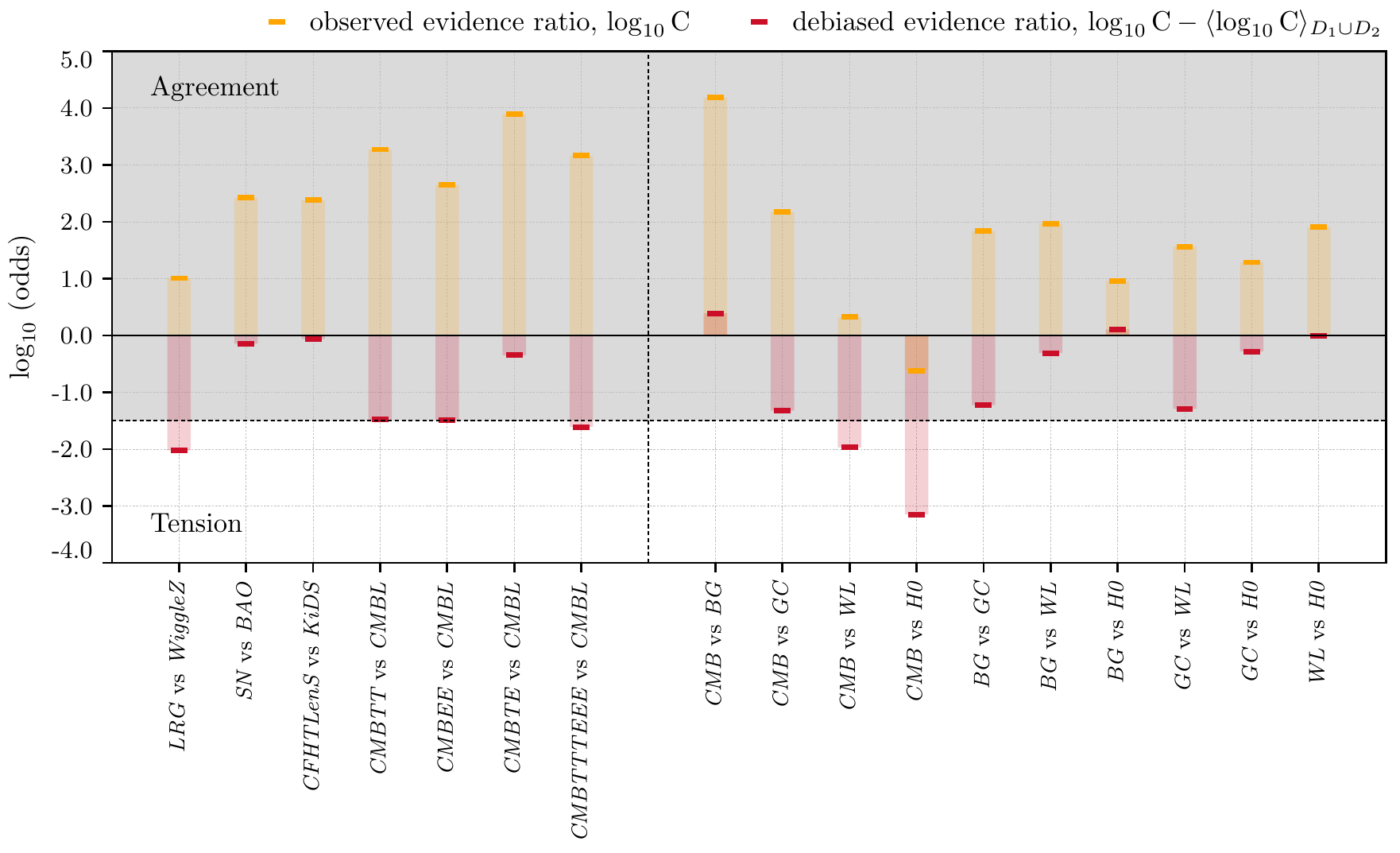}	
\caption{ \label{Fig:EvidenceRatio}
The evidence ratio estimator applied to different data set couples.
We show the nominal observed value of the evidence ratio test and its debiased value. 
Notice that for most of the data sets the bias in the evidence ratio estimator is as large as its observed value.
The darker shade indicates results would not be considered statistically significant on the Jeffreys' scale.
}
\end{figure*}

\begin{figure*}[!ht]
\centering
\includegraphics[width=\textwidth]{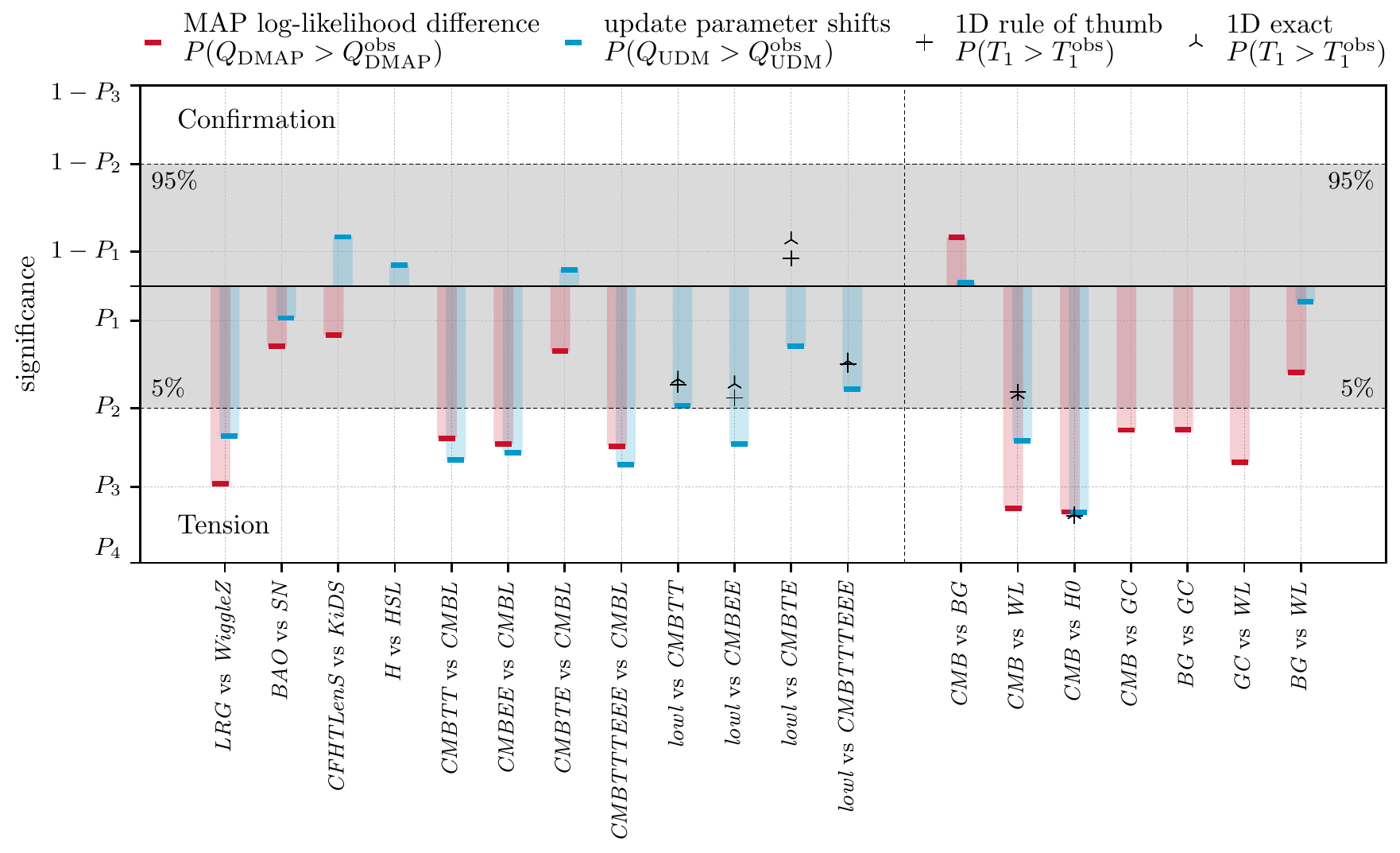}	
\caption{ \label{Fig:StatisticalSignificance}
The statistical significance of different CDEs for various data set couples: 
the difference in log-likelihood at maximum posterior (MAP),  $Q_{\rm DMAP}$
from Eq.~(\ref{Eq:DMAPsummary}), the update parameter shifts test, $Q_{\rm UDM}$
from Eq.~(\ref{Eq:UDMsummary}),
the exact 1D parameter shifts, $T_1$ from
Eq.~(\ref{Eq:T1twotailed}), and the ``rule of thumb difference in mean'', as the Gaussian approximation of $T_1$
from Eq.~(\ref{Eq:1DDifferenceMean}).
Different colors indicate different tests, as shown in legend. 
The labels report different levels of statistical significance: $P_1\equiv 32\%$, $P_2\equiv 5\%$, $P_3\equiv 0.3\%$, $P_4\equiv 0.007\%$.
Values that are identified as failure modes of one of the estimators are not shown in figure.
The darker shade indicates results that are not statistically significant.
}
\end{figure*}

In this section we present the application of the ratio of likelihoods at maximum posterior estimator  $Q_{\rm DMAP}$, introduced in Sec.~\ref{Sec:EvidenceRatio}, and discuss its relationship with the evidence ratio.

The practical challenges in computing the $Q_{\rm DMAP}$ estimator are the same as the maximum posterior goodness of fit and are mitigated in the same way that was discussed in the previous section.
The only difference is that, in the previous section, errors on $N_{\rm eff}$ had a small effect for all data sets that have a large number of data points. In this context, it is crucial to properly identify parameters, as the number of considered data points drops out of degrees of freedom counting, as shown in Sec.~\ref{Sec:EvidenceRatio}.
As we show in App.~\ref{App:Tables}, see Table~\ref{Table:EvidenceRatioResults}, the number of effective parameters for the single and joint data sets agrees well with physical intuition and that their difference appropriately reflects the number of parameters that both data sets are measuring.

Similarly to the previous section we cannot use the \data{lowl} and \data{HSL} data sets as their likelihood is non-Gaussian in the data. In addition we cannot apply this test to data set couples that are correlated and we have to exclude the comparison of the primary CMB spectra.

Before turning to $Q_{\rm DMAP}$ we apply the evidence ratio test to several data couples, as shown in Fig.~\ref{Fig:EvidenceRatio}, and subtract its bias, as computed within the GLM.
The evidence is estimated with the Gaussian approximation to the MCMC posterior, as discussed in App.~\ref{App:GaussianApproximationMCMC}, and its bias is computed using the statistics of that approximation.

The first noteworthy result that is shown in Fig.~\ref{Fig:EvidenceRatio} is that the observed value of the evidence ratio is usually of the same order of the bias in the evidence ratio.
This bias does also depend on the data set involved in the comparison and has to be subtracted case by case.
This shows the limitations of the evidence ratio test judged on the Jeffreys' scale. The results is usually so biased that the observed value alone cannot be used
to judge agreement or disagreement.

On the other hand, in Fig.~\ref{Fig:StatisticalSignificance}, we show the statistical significance of the $Q_{\rm DMAP}$ estimator.
The reported results confirms the picture that comes from the debiased evidence ratio while providing an estimate of statistical significance.
The qualitative agreement between the two is due to the fact that, when parameter space directions are either completely constrained by the prior or the data $Q_{\rm DMAP}$ is distributed as the evidence apart for additive factors that do not depend on the data realization and drop out of the
statistical significance.

We first consider the internal compatibility of data within the set families.
As we can see the \data{SN} and \data{BAO} data sets agree as well as the \data{CFHTLenS} and \data{KiDS} data sets making the \data{BG} and \data{WL} families internally consistent.
The \data{LRG} and \data{WiggleZ} data sets, on the other hand, show a marked indication of disagreement.
This is not surprising considering the indication of confirmation bias in the \data{LRG} data set and points toward a significant difference in parameters between the two probes.
This difference is not signaled by the ``rule of thumb difference in mean'', when applied to the $\Omega_m$ and $\Omega_b$ parameters, pointing toward a correlated shift in parameters.
Notice that, in this case, the bias in the evidence ratio is larger than the observed value. If we were to look the the latter and judge its value of the Jeffreys' scale we would draw the wrong conclusion that the two data sets agree.

Other interesting results concern the internal consistency of the \data{CMB} family.
The \data{CMBTT} and \data{CMBEE} data sets do not agree with \data{CMBL} at about a 5\% probability to exceed. For both data sets, this is roughly the same statistical significance of the deviation of the amplitude of the lensing parameter, $A_L$, from one, as reported in~\cite{Ade:2015xua,Motloch:2018pjy}.
While \data{CMBTE} and \data{CMBL} agree the joint result of \data{CMBTTTEEE} and \data{CMBL} is dominated by the tension in the temperature spectrum, consistently with the results in~\cite{Ade:2015xua,Motloch:2018pjy}.
Notice that the evidence ratio result obtained with the Gaussian approximation to \data{CMBTT} and \data{CMBL} agrees very well with the result of numerical integration shown in~\cite{Raveri:2015maa}.

We next apply the evidence ratio test to understand the compatibility of different families of physical probes.

As we can see in Fig.~\ref{Fig:StatisticalSignificance} the \data{CMB} family agrees well with the \data{BG} family but disagrees with the other three families of data sets that we consider.
The disagreement between \data{CMB} and \data{GC} families can be understood considering the indication of a confirmation bias in the \data{LRG} data set.
The statistical significance of the disagreement between these two probes roughly matches the statistical significance of confirmation in the \data{LRG} data set, pointing toward the hypothesis that the latter data set might be confirmation biased around parameter values that are not the \data{CMB} ones. 
The \data{CMB} data set also shows high statistically significant indications of tensions with the \data{WL} and \data{H0} data sets.
The tension with Hubble constant measurements is known and we recover $0.088\,\%$ probability to exceed compared with the ``rule of thumb difference in mean" result applied to $H_0$ of $0.073\%$ and the exact 1D shift that results in $0.078\%$.
The \data{WL} result is also known but has been usually evaluated using the full scale measurements of weak lensing, including scales that are influenced by the non-linear evolution of cosmological perturbations. Here we show that this tension persists and remains statistically significant, specifically at  $0.1\%$ probability to exceed, when restricting to linear scales.
Notice that the evidence ratio between the \data{CMB} and \data{H0} data set is the only one that is found negative. Still interpreting at face value this ratio on the Jeffreys' scale would lead to the incorrect conclusion that the tension is not significant.

The other data sets families considered generally agree. From a physical standpoint we know that they should since they are either measuring different parameters or weakly measuring the same parameters. This aspect is properly recovered and none of them are found to be in tension or confirmation biased at relevant statistical significance.
The only exception is the test applied to the \data{BG} and \data{WL} data sets against the \data{GC} data set.
The first is in tension with the latter as a consequence of its agreement with \data{CMB} at about the same statistical significance.
The second one is in tension with the latter due to the fact that both data sets have some problem at the goodness of fit level. Their combination is not surprisingly signaling disagreement of some sort.

\subsection{Parameter differences} \label{Sec:ResParameters}
In this section we present the application of the parameter shift CDE discussed in Sec.~\ref{Sec:ParametersQuad}.

The challenges in applying this CDEs to real data are profoundly different than the ones that we discussed in the previous sections.
This allows for a larger degree of complementarity between tests and ensures the robustness of conclusions against possible contamination from non-Gaussianities and other estimate problems.

In the following we  only use parameter difference estimator $Q_{\rm UDM}$ using  Eq.~(\ref{Eq:UDMQuadraticForm}), which is defined through the parameter update when combining two
data sets.
Parameter difference estimators of the form $Q_{\rm DM}$ using Eq.~(\ref{Eq:DMQuadraticForm}) have problems that are difficult to overcome in practical applications.
In case of uninformative flat priors any such test would be ill posed for directions that are unconstrained by one of the data sets. 
If we consider Gaussian priors, then the $Q_{\rm DM}$ itself can be formally defined.  However, noise in the determination of the covariances of the two experiments, due to MCMC sampling, makes it difficult to disentangle prior constrained and data constrained directions.
In applying it to the data we find this estimator to be unreliable and  numerical noise dominated for a wide variety of algorithms used for the estimate.

Aside from numerical issues, differences in parameters update also have the clear advantage that corrections due to non-Gaussianities are mitigated if the posterior of the most constraining data set is Gaussian.  In our cosmological applications CMB data play this role since parameter posteriors are nearly Gaussian for all $\Lambda$CDM parameters. 
If the second data set has a non-Gaussian posterior, a direct parameter difference would misestimate significance if the mean of the first set lay in the tail of the second set.  
For the parameter update, GLM is effectively applied around the mean of the first set by replacing the non-Gaussian posterior of the second set with a Gaussian approximation locally  around that point.

To minimize numerical noise in the $Q_{\rm UDM} $ estimates we use the Karhunen-Loeve (KL) decomposition of the two covariances that are involved.  Recall that 
to compute the observed value of the update parameter shift we need to evaluate:
\begin{align}
Q_{\rm UDM} \equiv (\Delta\bar{\theta}_{U} )^T \,(\mathcal{C}_{p1} -\mathcal{C}_{p\joint})^{-1}\,(\Delta\bar{\theta}_{U}  )  \,. 
\label{eqn:QUDM}
\end{align}
The second data set can only add information on top of the first data set so that $(\mathcal{C}_{p1} -\mathcal{C}_{p\joint})$ has to be positive definite in the absence of numerical noise. In the presence of numerical noise, it is better to first transform to the KL
basis since it is mutually orthogonal in the metrics defined by $\mathcal{C}_{p1}$ and $\mathcal{C}_{p\joint}$.   
We  solve the generalized eigenvalue problem to find the KL modes, $\phi^{a}$, of the two covariances:
\begin{align}
\sum_{\nu} \mathcal{C}_{p1}^{\mu\nu} \, \phi_{\nu}^{\,\,a} = \lambda^{a} \sum_{\nu} \mathcal{C}_{p\joint}^{\mu\nu} \, \phi_{\nu}^{\,\,a} \,.
\end{align}
Here the eigenmodes are defined to be orthonormal in the $\mathcal{C}_{p\joint}$ metric
\begin{equation}
\sum_{\mu\nu} \phi_{\mu}^{\,\,a}\mathcal{C}_{p\joint}^{\mu\nu}  \phi_{\nu}^{\,\,b} = \delta^{ab},
\end{equation}
and since they are orthogonal in the $\mathcal{C}_{p1}$ metric, but with variance $\lambda^a \delta^{ab}$, the KL basis provides
 linear combinations of the parameters that are mutually independent and ordered by the improvement in the variance of $\joint$ over
 $1$. 
If we now define the linear combination of parameter differences in the KL basis as
\begin{equation}
\Delta p^a = \sum_\mu  \phi_{\mu}^{\,\,a}\Delta\bar{\theta}_{U}^\mu
\end{equation}
we obtain
\begin{equation}
Q_{\rm UDM} \equiv \sum_{a=1}^{N_{\rm KL}} \frac{ (\Delta p^a)^2}{\lambda^a -1 }.
\end{equation} 
While this transformation, when $N_{\rm KL}$ is the full set of KL modes, gives exactly the same value as
Eq.~(\ref{eqn:QUDM}), it also highlights the problem of numerical noise.   If $\joint$ does not improve over
$1$ substantially in a given mode, then $\lambda^a \approx 1$ and numerical noise in the estimation of covariances create large
errors in $Q_{\rm UDM}$.
The KL decomposition allows us to place a well defined lower cutoff on this improvement in order to remove unwanted numerical noise from the estimator.
In practical applications there is a hierarchy of KL modes so that noise and data modes are well separated in the spectrum. We use a simple algorithm to find this separation point and define the optimal cutoff for each data set combinations.
To minimize numerical noise in the $Q_{\rm UDM} $ estimates we also notice that it is preferable to use the mean of the parameters in the test rather than the best fit parameters even though they are the same in the GLM.

We find that for parameter distributions that are well approximated by Gaussian distributions the cutoff is usually in the range of $5\%$ while it can be as large as $15\%$ in case of non-Gaussian posteriors.
In all cases we limit the cutoff to be between $2\%$ and $20\%$ and we cannot extend it to zero otherwise the estimator will be noise dominated.
Notice that this prescription also effectively defines 
\begin{equation}
\langle Q_{\rm UDM}  \rangle = N_{\rm KL} \,,
\end{equation}
and hence $Q_{\rm UDM}$ is chi-squared distributed with $N_{\rm KL}$ degrees of
freedom.

With this technique the estimator is stable but is left with one case where the statistic returns a null result. 
When we are combining a data set that is very constraining with a data set that is very weakly constraining the improvement in the KL modes might be below the threshold that separates data dominated modes and noise dominated modes. 
In this case the value of $Q_{\rm UDM}$ will be zero and it distributed as a $\chi^2$ with zero degrees of freedom, i.e. zero for all data
realizations. This simply means that while there may be a true, but tiny, parameter shift, it is too small to measure.  In this case, the procedure correctly returns that the answer that neither tension nor confirmation bias can be detected.  

We start by applying the update difference in mean to assess the consistency of data sets families and we report the results in Fig.~\ref{Fig:StatisticalSignificance}.

As we can see the disagreement between the \data{LRG} and \data{WiggleZ} data sets, at parameter level, is confirmed to be statistically significant, as we found in the previous section. 
The statistical significance of this result is, however, slightly lower than what is reported by the likelihood at maximum posterior test.
This effect can be attributed to the different sensitivity of the two estimators to the effect of non-Gaussianities in the parameter posteriors.

The parameter update results further confirm the internal consistency of the \data{BG} and \data{WL} families, as found in the previous section.
On the other hand, we can extend here the study of internal consistency of families of probes to include \data{H} and \data{HSL} measurements. While the latter has a likelihood that is non-Gaussian at the data level, at parameter level it can be well approximated by a Gaussian distribution.
As we can see the two data sets agree on the determination of the Hubble constant while not showing indications of tensions nor confirmation.

Similarly we can here extend the study of the \data{CMB} internal consistency, even though the \data{lowl} likelihood is non-Gaussian at the data level.
The update parameter shift test confirms the tension between \data{CMBTT}, \data{CMBEE} and \data{CMBL} and the agreement between \data{CMBTE} and \data{CMBL}, at about the same statistical significance that was found in the previous section.

If we now consider the same set of comparisons, with the addition of the CMB large angular scale multipoles, we see that the agreement between the primary CMB spectra and \data{CMBL} improves to the point that it is not statistically significant.
This picture is consistent with the results of the update parameter shift test applied between the \data{lowl} data and the primary CMB spectra.
As we can see from Fig.~\ref{Fig:StatisticalSignificance} all the four results are on the tension side and exceed $95\%$ C.L. for the \data{CMBEE} spectrum.
These are also in qualitative agreement with the the ``rule of thumb'' and 1D shift when applied to the $\tau$ parameter.
The tension is reported to be slightly larger because the direction that is selected by the KL decomposition takes into account degeneracies with other cosmological parameters.

The discrepancy between these three probes can be physically understood because at fixed $A_s e^{-2\tau}$, lowering $\tau$ reduces $A_s$ and hence reduce the gravitational lensing potential and the smoothing of the CMB peaks.
At high multipoles the CMB measurements of Planck have enough precision to be sensitive to gravitational lensing, hence other parameters shift to compensate for the decreased smoothing of the peaks. 
This is achieved by increasing $\Omega_m h^2$ and $A_s e^{-2\tau}$, while reducing $n_s$ and $\Omega_b h^2$, as discussed in~\cite{Aghanim:2016sns}.
The best fit solution to the \data{lowl}+\data{CMBTT} has known oscillatory residuals at high multipoles~\cite{Aghanim:2016sns} because of the lack of power at large angular scales.
Without the \data{lowl} data set these oscillatory residuals can be fit by raising $\tau$ that is balanced by raising also $A_s$ and $\Omega_m h^2$ that overall give a larger CMB lensing signal that is in conflict with lensing reconstruction of the \data{CMBL} data set.
This tension can then be isolated by adding a new parameter that describes the amplitude of the lensing of the CMB, $A_L$, that allows to fit the oscillatory residuals in the primary spectra and is found to be deviating form unity at about the statistical significance of the tensions that we report here. 

We can now proceed to the application of the update parameter shift test to different data sets families.
These results are largely in agreement with the ones reported in the previous section with some noticeable differences.
As shown in App.~\ref{App:Tables} these results do not depend strongly on the inclusion of the \data{lowl} data set that leaves them largely unchanged. 

While the tension between \data{CMB} and \data{H0} is confirmed and in good agreement with the benchmark results,
specifically at $0.087\%$ agreement probability, the tension between the \data{CMB} and \data{WL} data sets is markedly lower than the $Q_{\rm DMAP}$ result, specifically at $1.6\%$ agreement probability.
This is expected since the \data{WL} data set does show a non-Gaussian posterior with slowly decaying tails.
Still this tension is noticeably higher with respect to the ``rule of thumb'' estimate applied to the $S_8$ parameter which yields $7.1\%$ agreement probability and the exact 1D shift that takes into account the non-Gaussianity of the posterior and results in $6.7\%$ agreement probability.

In this case the $Q_{\rm UDM}$ test is indicating, through the number of degrees of freedom, that this tension is evaluated along one parameter space direction, $\langle Q_{\rm UDM}\rangle = 1$. This direction is built to be the optimal one for both data sets.  
The $S_8$ parameter, in turn, is not exactly describing the amplitude of the lensing signal, at the redshifts of the combined \data{WL} surveys that we are considering, and is not the best constrained parameter.
We find that, for the \data{WL} data set, $\sigma_8 \Omega_m^{0.7}$ is better constrained and the ``rule of thumb'' test is signaling a tension similar to that of the $Q_{\rm UDM}$ estimator.

Finally we can easily see an example of the null result mode of the estimator by looking at data sets combinations involving the \data{GC} data set.
This data set is very weakly constraining, when compared with the \data{CMB} data set, so that its improvement on the parameter constraints cannot be distinguished from numerical noise.
Furthermore the \data{GC} data set is very weakly constraining along the parameter space directions that are constrained by the \data{WL} data set so that the result is again dominated by noise and our KL procedure properly identifies this as a null update result.

%
\section{Conclusions} \label{Sec:Conclusions}
%
We studied statistical estimators of concordance and discordance (CDEs) between cosmological probes and applied them to state of the art cosmological data sets.

We discussed the likelihood at maximum posterior as a measure of the goodness of fit. Unlike the maximum likelihood, this quantity depends on the prior on cosmological parameters and allows to disentangle parameter space directions that are constrained by the data and by the prior.
This disentanglement provides a fair degree-of-freedom counting when performing the goodness of fit test.

We studied the distribution of the evidence ratio test of data set compatibility over the space of data realizations.
This allowed us to uncover the fact that the evidence ratio is usually biased toward agreement and that, in practical applications, this bias is as large as the observed value, making the Jeffreys' scale unreliable as an indicator of  agreement or disagreement.
We then defined a similar estimator based on the ratio of likelihoods at maximum posterior that allows for an assessment of statistical significance of the reported results.
While being equivalent to the evidence ratio in the limiting cases where parameter space directions are completely constrained by either the data or the prior, this estimator is significantly easier to apply.

We investigated the statistics of parameter shifts developing methods that work in arbitrary number of dimensions.
These estimators optimally weight the parameter shifts and mitigate the fact that tensions might not be identifiable at the single parameter level because
they are hidden by the process of marginalization over a high dimensional parameter space.  We introduce a robust regularization scheme based on the
Karhunen-Loeve decomposition which identifies and discounts the small parameter shifts due to sampling noise in MCMC posteriors.   

When applying these estimators to cosmological data we find several noteworthy results.
As a benchmark for the estimators we recover the known result regarding tensions between the Planck measurements of the CMB spectra and local measurements of the Hubble constant and the amplitude of the galaxy weak lensing signal.
Concerning the latter, we find that, when considering the Canada-France-Hawaii Telescope Lensing Survey and the Kilo Degree Survey on large linear scales the statistical significance of the disagreement with CMB measurements is between $98.4\%$ and $99.9\%$.
This is somewhat higher than is estimated by looking at then posterior of the $S_8\equiv \sigma_8 \Omega_m^{0.5}$ parameter alone as we optimally weight all parameter space directions.

We investigated the consistency of CMB measurements of the Planck satellite, establishing a set of results that allow to prioritize the analysis of the next release of the Planck data.
In particular we find that: the CMB TE cross correlation is a bad fit and that seems to be related to the presence of residual, frequency dependent foregrounds;
the discrepancy between the CMB TT spectrum and its lensing reconstruction is also present in the E-mode spectrum at about the same statistical significance;
the measurements of the large angular scales multipoles, $\ell < 30$, are in tension with the small scale temperature and E-mode spectra at about $95\%$ probability.

Moreover we find CMB results to be in tension with probes of the clustering of galaxies. This disagreement can be likely attributed to the SDSS LRG DR4 survey being slightly confirmation biased toward a different set of cosmological parameters.

The work toward understanding the consistency of present cosmological probes and preparing for the analysis of the next generation of probes is far from complete. 
Future efforts in these directions include the generalization of the techniques presented in this paper to consider non-Gaussian corrections.
Moreover we need to develop statistical estimators that work on more than two data sets at the time, allowing to compute the joint distribution of multiple tests.
These will allow us to understand the global consistency of the $\Lambda$CDM model with a large and diverse set of experimental data.

Finally these tests should be applied as we gather new and more precise cosmological data sets to make sure that inconsistencies due to systematic effects or incomplete modeling of cosmological observables are identified and corrected and that discrepancies due to new physical phenomena are promptly found.

\acknowledgments
We thank
Chihway Chang,
Tom Crawford,
Pavel Motloch,
Sam Passaglia and 
Kimmy Wu 
for useful comments.
MR and WH are supported by U.S.~Dept.~of Energy contract DE-FG02-13ER41958.  
WH was further supported by NASA ATP NNX15AK22G and the Simons Foundation.
Computing resources were provided by the University of Chicago Research Computing Center through the Kavli Institute for Cosmological Physics at the University of Chicago. 
\appendix
\section{Quadratic forms in Gaussian random variables} \label{App:QuadraticForms}
In this appendix we briefly outline how to practically deal with the statistics of the many quadratic forms that appear in the main text.
This material is mostly taken from~\cite{mathai1992quadratic} and reproduced here to ease the comprehension of the main text. 

A quadratic form in the $p$ dimensional random Gaussian variable $X$ is defined by:
\begin{align} \label{EqApp:QuadraticForm}
Q = X^T A X \,\,; \hspace{0.5cm} X\sim \mathcal{N}_p( x; \mu,\Sigma) \,.
\end{align}
The first two moments of the quadratic form are:
\begin{align} \label{EqApp:QuadraticFormMomenta}
\langle Q\rangle_X =\,& {\rm tr} [A\Sigma]+\mu^T A \mu \,, \nonumber \\
{\rm Var}(Q) =\,& 2\,{\rm tr} [(A\Sigma)^2] +4\mu^TA \Sigma A\mu \,.
\end{align}
In the following we only consider the case of central quadratic forms $\langle X \rangle=\mu = 0$. We find that all distributions in the main text satisfy this requirement. For the generalization of the following results to the case where $\mu \neq 0$ we refer the reader to~\cite{mathai1992quadratic}.

Over the subspace where $\Sigma$ is invertible, $Q$ admits a decomposition of the form:
\begin{align} \label{EqApp:QuadraticFormDecomposition}
Q=X^TAX = \sum_{j=1}^{p} \lambda_j U_j^2 \,,
\end{align}
where $\lambda = {\rm eigenval}\left( A \Sigma \right)$, $P = {\rm eigenvec}\left( A \Sigma \right)$ and
\begin{align}
U =\,& P^T \Sigma^{-1/2}( x-\mu) \sim \mathcal{N}_p(x;0,\mathbb{I}) \,.
\end{align}
Given that $U_j$ is a normally distributed variable, $U_j^2$ is a $\chi^2(1)$ variable, and
so $Q$ is in general distributed as the sum of scaled  $\chi^2(1)$ variables which are themselves  known as Gamma distributed variables.

If $A$ is any projection of $\Sigma^{-1}$, i.e. $A = \mathbb{P}^T \Sigma^{-1} \mathbb{P}$ where $\mathbb{P}^2=\mathbb{P}$, then all the eigenvalues $\lambda_j \in 0,1$ and $Q$ would be the sum of independent $\chi^2(1)$  variables, $Q \sim \chi^2(r)$, with $r={\rm rank} (\mathbb{P})$ degrees of freedom.  This includes the trivial case where $A= \Sigma^{-1}$.

More generally, if all the eigenvalues $\lambda_j \ge 0$  then 
analytic expressions for the probability density of $Q$ exist~\cite{mathai1992quadratic} and probabilities can be computed with dedicated algorithms~\cite{CompQuadForms} once the eigenvalues of $A \Sigma$ are obtained.

Alternately, the distribution of $Q$ can  be approximated by that of a chi squared variable matching some of the moments of $Q$. We refer to these as Patnaiks' type approximations~\cite{10.2307/2332149}.
The first approximation matches the mean to a (single) chi square distribution:
\begin{align} \label{Eq:Patnaiks1}
Q = \sum_j \lambda_j X_j^2 \simeq \chi^2(  {\rm tr}[A\Sigma] ) \,,
\end{align}
where $\simeq$ stands for ``approximately distributed as''.
The second approximation matches the mean and variance to that of (single) Gamma distribution:
\begin{align} \label{Eq:Patnaiks2}
Q = \sum_j \lambda_j X_j^2 \simeq c \chi^2( \nu ) \,,
\end{align}
where:
\begin{align}
& c \equiv \sum_j \lambda_j^2/ {\rm tr}[A\Sigma] \,,\nonumber \\
& \nu \equiv ({\rm tr}[A\Sigma])^2 / \sum_j \lambda_j^2 \,.
\end{align}
Notice that in both approximations the number of degrees of freedom of the (scaled) chi square distribution is usually not integer.

We shall use the first approximation in practice, matching only the mean.  
When $0\leq \lambda \leq 1$, this approximation is conservative as the second approximation and the full distribution have smaller variance.
These types of approximations are usually relevant over parameter space directions that are partially constrained by the prior, where the full posterior of the data set that we consider is usually highly non-Gaussian.
Underestimating their contribution to the statistical significance of the reported results is then a mitigation strategy against non-Gaussianities.

\section{Proofs of Section~\ref{Sec:GoF}} \label{App:ProofsGoF}
In this appendix we provide the proofs for the results contained in Sec.~\ref{Sec:GoF} as a worked example of how to use the GLM in practice.

We first consider the maximum likelihood:
\begin{align}
-2 \ln \mathcal{L}_{\rm max} =&\, X^T (\mathbb{I}-\mathbb{P})^T \Sigma^{-1} (\mathbb{I}-\mathbb{P}) X \nonumber \\
&\, +d\ln(2\pi) +\ln(|\Sigma|) \,,
\end{align}
where, here and below, $X \equiv x -\hat m$ with $\hat m=m_\Pi$ for convenience, involves the component of the
data vector that is in the complement of the model space $ (\mathbb{I}-\mathbb{P}) X$.
For any prior, this data vector is distributed as
\begin{equation}
 (\mathbb{I}-\mathbb{P}) X \sim {\cal N}_{r}((\mathbb{I}-\mathbb{P}) X;0, (\mathbb{I}-\mathbb{P})\Sigma
 (\mathbb{I} -\mathbb{P})^T) \,,
 \end{equation}
 where $r = {\rm rank}(\mathbb{I}-\mathbb{P})$, since the projector nulls the part of the data draw covariance that lives on parameter space. 
The data dependent piece of the maximum likelihood statistic  contains
\begin{equation}
Q_{\rm ML} = [ (\mathbb{I}-\mathbb{P}) X] ^T \Sigma^{-1} [(\mathbb{I}-\mathbb{P}) X ] \,,
\end{equation}
which is a quadratic form for this data vector. Considering now the results of the previous section, the statistics of $Q_{\rm ML}$ is
 determined by the eigenvalues
 \begin{align}
\lambda =&\, {\rm eigenval}\left[ \Sigma^{-1} (\mathbb{I}-\mathbb{P})\Sigma
 (\mathbb{I} -\mathbb{P})^T \right] \nonumber \\
=&\, {\rm eigenval}(\mathbb{I}-\mathbb{P}) \,,
\end{align}
which implies  $Q_{\rm ML} \sim \chi^2(r)$.
If we assume that the model has $N$ relevant parameters then $\mathbb{P}$ projects the
data vector onto an $N$ dimensional subspace and therefore its complement has $r= d-N$. 

We now turn to the distribution of the evidence.
For uninformative flat priors, the evidence quadratic form is identical to the maximum likelihood
quadratic form and is therefore also distributed as $\chi^2(d-N)$.
In case of delta priors, the evidence quadratic form in data space is
\begin{align} \label{Eq:AppDeltaEvGoF}
Q_\mathcal{E} = X^T\Sigma^{-1}X \,,
\end{align}
where $X$ is normally distributed with covariance $\Sigma$.  Thus
$Q_\mathcal{E}$ is chi squared distributed with full rank $\chi^2(d)$.

In case of Gaussian priors the quadratic form defined by the evidence becomes:
\begin{align} \label{Eq:AppGaussEvGoF}
Q_\mathcal{E} = X^T \Sigma_0^{-1} X \,,
\end{align}
where $X$ is normally distributed with covariance $\Sigma_0= \Sigma +M\mathcal{C}_\Pi M^T$.
Thus $Q_\mathcal{E}$ is again distributed as $\chi^2(d)$.

We next derive the distribution of the likelihood at maximum posterior for different prior choices.
In case of delta priors we have:
\begin{align}
-2 \ln \mathcal{L}(\theta_p) =&\, -2 \ln \mathcal{L}(\theta_{\Pi}) \nonumber \\
=&\, X^T\Sigma^{-1}X +d\ln(2\pi) +\ln(|\Sigma|) \,,
\end{align}
that, up to a constant, defines a quadratic form in data space:
\begin{align}
Q_{\rm MAP}  =&\, X^T\Sigma^{-1}X \,,
\end{align}
that is distributed as $\chi^2(d)$.

In case of uninformative flat priors the likelihood at maximum posterior is just the maximum likelihood and so $Q_{\rm MAP} = Q_{\rm ML}$.

Gaussian priors stand in between these cases.
The data dependent part, $Q_{\rm MAP} $, of the likelihood at maximum posterior is given by:
\begin{align} \label{Eq:AppMaxPostGauss}
Q_{\rm MAP}  =&\, -2 \ln \mathcal{L}(\theta_p) -d\ln(2\pi) -\ln(|\Sigma|) \nonumber \\
=&\, X^T\bigg[ (\mathbb{I}-\mathbb{P})^T \Sigma^{-1} (\mathbb{I}-\mathbb{P}) \nonumber \\
&\, +\tilde{M}^T\mathcal{C}_{\Pi}^{-1} \mathcal{C}_p \mathcal{C}^{-1} \mathcal{C}_p \mathcal{C}_\Pi^{-1} \tilde{M} \bigg]X \,.
\end{align} 
This quadratic form has $d-N$ eigenvalues with $\lambda_i=1$ together with the $N$ eigenvalues of $\mathcal{C}_{\Pi}^{-1} \mathcal{C}_p$ that are 
bounded as $0 \le \lambda_i \le 1$.
From this set of eigenvalues the exact distribution can be computed or approximated as in App.~\ref{App:QuadraticForms}.
If we take the first approximation, Eq.~(\ref{Eq:Patnaiks1}), that matches the mean, then $Q_{\rm MAP}  \simeq \chi^2( d-N+{\rm tr}[\mathcal{C}_{\Pi}^{-1} \mathcal{C}_p]) = \chi^2( d-N_{\rm eff})$.
This approximation is exact for all parameter space directions that are data dominated $\mathcal{C}_{\Pi}^{-1} \mathcal{C}_p \rightarrow 0$ or completely prior dominated $\mathcal{C}_{\Pi}^{-1} \mathcal{C}_p \rightarrow 1$ and approximated for cases in between.

As long as the number of partially constrained directions in parameter space remains small compared with the
total number of degrees of freedom, the approximation works very well.
When the number of partially constrained directions is large,  Eq.~(\ref{Eq:Patnaiks1}) systematically underestimates the statistical significance of results. In such cases, however, it is likely that the distributions that are being considered are highly non-Gaussian so that results should be interpreted with caution anyway.

\section{Proofs of Section~\ref{Sec:EvidenceRatio}} \label{App:ProofsEvR}
In this appendix we report the proofs of the results contained in Sec.~\ref{Sec:EvidenceRatio}.

We start by discussing the statistics of the ratio between the maximum likelihoods of two experiments and their joint maximum likelihood.
In data space this can be written as:
\begin{equation} 
-2\Delta \- \ln \mathcal{L}_{\rm max} \equiv -2\ln \mathcal{L}_{\rm max}^{\joint} +2\ln \mathcal{L}_{\rm max}^{1} +2\ln \mathcal{L}_{\rm max}^{2} \,. 
\end{equation}
Since we assumed that the two data sets are uncorrelated, the data independent part cancels so that $-2\Delta \ln \mathcal{L}_{\rm max}$ defines a quadratic form in data space:
\begin{align} \label{Eq:AppMLRatio}
Q_{\rm DML}=&\, X_{\joint}^T (\mathbb{I}_{\joint}-\mathbb{P}_{\joint})^T \Sigma^{-1}_{\joint}  (\mathbb{I}_{\joint}-\mathbb{P}_{\joint}) X_{\joint} \nonumber \\
&\, - X_{1}^T (\mathbb{I}_{1}-\mathbb{P}_{1})^T \Sigma^{-1}_{1}  (\mathbb{I}_{1}-\mathbb{P}_{1}) X_{1} \nonumber \\
&\, - X_{2}^T (\mathbb{I}_{2}-\mathbb{P}_{2})^T \Sigma^{-1}_{2}  (\mathbb{I}_{2}-\mathbb{P}_{2}) X_{2} \,,
\end{align}
where $X_{\joint} \equiv x_{\joint}-\hat{m}_{\joint}$, $X_{1} \equiv x_{1}-\hat{m}_{1}$, $X_{2} \equiv x_{2}-\hat{m}_{2}$ and we assumed that the two data sets are independent so that $\Sigma_{\joint} = {\rm diag}(\Sigma_1,\Sigma_2)$.

The projector $\mathbb{P}_{\joint}$ takes data realizations of the joint data set $(x_1,x_2)$ and projects them on the model tangent space. 
It can be explicitly written as:
\begin{align}
\mathbb{P}_{\joint} =&\, M_{\joint}(M_{\joint}^T\Sigma_{\joint}^{-1}M_{\joint})^{-1}M_{\joint}^T \Sigma_{\joint}^{-1} \nonumber \\
=&\,
\left(\begin{array}{cc}
M_1\mathcal{C}_{\joint}M_1^T\Sigma^{-1}_{1} & M_1\mathcal{C}_{\joint}M_2^T\Sigma^{-1}_{2}  \\
M_2\mathcal{C}_{\joint}M_1^T\Sigma^{-1}_{1} & M_2\mathcal{C}_{\joint}M_2^T\Sigma^{-1}_{2}  \\
\end{array}\right) \,,
\end{align}
to verify that it is a projector $\mathbb{P}_{\joint}^2 = \mathbb{P}_{\joint}$ and that it leaves the tangent space of the model invariant $\mathbb{P}_{\joint} M_{\joint} = M_{\joint}$.
Notice that the projector of the joint data set cannot be written as the direct sum of the two single projectors ${\rm diag}(\mathbb{P}_{1},\mathbb{P}_{2})$ but it can be shown by direct calculation that they commute:
\begin{align}
\mathbb{P}_{\joint} \,\, {\rm diag}(\mathbb{P}_{1},\mathbb{P}_{2}) = {\rm diag}(\mathbb{P}_{1},\mathbb{P}_{2}) \, \mathbb{P}_{\joint}  =  \mathbb{P}_{\joint}  \,.
\end{align}
This implies that the subspace that $\mathbb{P}_{\joint}$ spans is contained in the subspace that ${\rm diag}(\mathbb{P}_{1},\mathbb{P}_{2})$ spans, $\mathbb{P}_{\joint} \subset {\rm diag}(\mathbb{P}_{1},\mathbb{P}_{2})$.
Conversely ${\rm diag}( \mathbb{I} -\mathbb{P}_{1},  \mathbb{I} -\mathbb{P}_{2}) \subset \mathbb{I} -\mathbb{P}_{\joint}$.

Now, by noticing that $\hat{m}_{\joint}=(\hat{m}_1,\hat{m}_2)$, we can write Eq.~(\ref{Eq:AppMLRatio}) as:
\begin{align} \label{Eq:AppMLRatio2}
 Q_{\rm DML} & =
 X_{\joint}^T\Big[  (\mathbb{I}_{\joint}-\mathbb{P}_{\joint})^T \Sigma^{-1}_{\joint}  (\mathbb{I}_{\joint}-\mathbb{P}_{\joint})  \\
& \hspace{0.35cm} 
-
\tiny
\Big(\begin{array}{cc}
(\mathbb{I}_{1}-\mathbb{P}_{1})^T \Sigma^{-1}_{1}  (\mathbb{I}_{1}-\mathbb{P}_{1})  & \mathbb{O} \\
\mathbb{O}       &  (\mathbb{I}_{2}-\mathbb{P}_{2})^T \Sigma^{-1}_{2}  (\mathbb{I}_{2}-\mathbb{P}_{2})
\end{array}\Big)
\Big] X_{\joint} \,,\nonumber
\end{align}
where $X_{\joint}$ is distributed according to the evidence of the joint data set.

In the delta prior case the joint evidence is given by $\mathcal{E}_{\joint} = \mathcal{N}_{d_1+d_2}( x_{\joint}; \hat{m}_{\joint}, \Sigma_{\joint} )$ while the uninformative flat prior case is $\mathcal{E}_{\joint} = \mathcal{N}_{d_1+d_2}( x_{\joint}; \hat{m}_{\joint}, [(\mathbb{I}-\mathbb{P}_{\joint})^T\Sigma^{-1}_{\joint}(\mathbb{I}-\mathbb{P}_{\joint})]^{-1} )$ and the Gaussian case has $\mathcal{E}_{\joint} = \mathcal{N}_{d_\joint}( x_{\joint}; m_{\joint \Pi }, \Sigma_{\joint} +M_{\joint}\mathcal{C}_{\Pi}M_{\joint}^T )$. 

When we compute the eigenvalues of the product of the matrix defining the quadratic form in Eq.~(\ref{Eq:AppMLRatio2}) and the covariance of the joint data draws it is sufficient to notice that in all three cases the projector would null everything along the model joint tangent space and so would ${\rm diag}( \mathbb{I} -\mathbb{P}_{1},  \mathbb{I} -\mathbb{P}_{2})$ since it is contained in $\mathbb{I} -\mathbb{P}_{\joint}$.
We can then apply Theorem (5.1.6) in~\cite{mathai1992quadratic} to show that Eq.~(\ref{Eq:AppMLRatio2}) is distributed as:
\begin{align}
&\chi^2( {\rm rank}(\mathbb{I}-\mathbb{P}_{\joint}) -{\rm rank}(\mathbb{I}-\mathbb{P}_{1})  -{\rm rank}(\mathbb{I}-\mathbb{P}_{2}) ) \nonumber \\
&= \chi^2(d_{\joint}-N_{\joint} -d_1+N_1-d_2+N_2) \nonumber \\
&= \chi^2(N_1 +N_2 -N_{\joint}) \,,
\end{align}
where we used the fact that the rank of a block diagonal matrix is the sum of the ranks of the diagonal blocks, having allowed different data sets to have different irrelevant parameters and noticing that $d_{\joint}=d_1+d_2$.

The same result can be obtained by bringing Eq.~(\ref{Eq:AppMLRatio}) in parameter space to show that:
\begin{align}
Q_{\rm DML} = (\theta_{\rm ML}^2-\theta_{\rm ML}^1)^T(\mathcal{C}_1+\mathcal{C}_2)^{-1}(\theta_{\rm ML}^2-\theta_{\rm ML}^1) \,, 
\end{align}
and considering $\theta_{\rm ML}^1$ and $\theta_{\rm ML}^2$ to be drawn independently from a Gaussian distribution with covariance $\mathcal{C}_1$ and $\mathcal{C}_2$ respectively.

We now consider the distribution of the evidence ratio.
In case of delta priors the distribution is trivial:
\begin{align}
-2 \Delta \ln \mathcal{E} \equiv&\, -2 \ln \mathcal{E}_{\joint} +2 \ln \mathcal{E}_{1} +2 \ln \mathcal{E}_{2} \nonumber \\
=&\, -2 \ln \mathcal{L}_{\joint}(\theta_\Pi) +2 \ln \mathcal{L}_{1}(\theta_\Pi) +2 \ln \mathcal{L}_{2}(\theta_\Pi) \nonumber \\
=&\, 0 \,.
\end{align}
Notice that we assume that the prior is the same for the analysis of the joint data set and the single data sets.
If this is not the case and the prior is changed between the analysis of different data sets the distributions of this appendix would be more complicated and, in general, non-central.

In the uninformative flat prior case the distribution of the data dependent part of the evidence ratio follows that of the maximum likelihood
$Q_{{\rm D}\mathcal{E}}  = Q_{\rm DML}$.

In the Gaussian case the distribution is more complicated and can be written starting from:
\begin{align} \label{Eq:AppEvRGaussian1}
-2\Delta &\-\ln \mathcal{E} = \nonumber \\
&\, -2 \ln \mathcal{L}_{\joint}(\theta_p^{\joint}) +2 \ln \mathcal{L}_{1}(\theta_p^{1})+2 \ln \mathcal{L}_{2}(\theta_p^{2}) \nonumber \\
&\, -2\ln\frac{\Pi_{\joint}(\theta_p^{\joint})}{\Pi_{\joint}^{\rm max}} +2\ln\frac{\Pi_{1}(\theta_p^{1})}{\Pi_{1}^{\rm max}}+2\ln\frac{\Pi_{2}(\theta_p^{2})}{\Pi_{2}^{\rm max}} \nonumber \\
&\, +(N_1+N_2-N_{\joint}) \ln (2\pi) \nonumber \\
&\, +2 \ln \frac{V_{\Pi}^{\joint}}{V_\Pi^1V_\Pi^2} +\ln \frac{|\mathcal{C}_{p1}||\mathcal{C}_{p2}|}{|\mathcal{C}_{p\joint}|} \,.
\end{align}
The data dependent part of Eq.~(\ref{Eq:AppEvRGaussian1}) defines a quadratic form in data space:
\begin{align} \label{Eq:AppEvRGaussian}
Q_{{\rm D}\mathcal{E}} \equiv&\, X_{\joint}^TAX_{\joint}  \\
=&\,X_{\joint}^T\bigg[  (\Sigma_{\joint} +M_{\joint}\mathcal{C}_{\Pi}M_{\joint}^T)^{-1} \nonumber \\
& -
\tiny
\left(\begin{array}{cc}
(\Sigma_{1} +M_{1}\mathcal{C}_{\Pi}M_{1}^T)^{-1}   & \mathbb{O} \\
\mathbb{O}       &  (\Sigma_{2} +M_{2}\mathcal{C}_{\Pi}M_{2}^T)^{-1} 
\end{array}\right)
\bigg] X_{\joint} \,,
\nonumber 
\end{align}
where $X_{\joint} \equiv x_{\joint} -m_{\joint}(\theta_\Pi)$ and the covariance of the joint data draws is explicitly given by:
\begin{align} \label{Eq:AppCovarianceDataGaussian}
\Sigma_{\joint}& +M_{\joint}\mathcal{C}_{\Pi}M_{\joint}^T =  \\
&=
\left(\begin{array}{cc}
    \Sigma_1       & \mathbb{O} \\
    \mathbb{O}       & \Sigma_2 
\end{array}\right)
+
\left(\begin{array}{cc}
    M_1 \mathcal{C}_\Pi M_1^T       & M_2 \mathcal{C}_\Pi M_1^T \\
    M_1 \mathcal{C}_\Pi M_2^T       & M_2 \mathcal{C}_\Pi M_2^T
\end{array}\right) \,. \nonumber
\end{align}
By direct computation of the product between the two matrices, 
\begin{align}
\lambda =&\, {\rm eigenval}\left[ A (\Sigma_{\joint} +M_{\joint}\mathcal{C}_{\Pi}M_{\joint}^T) \right] \nonumber \\
=&\, \pm \sqrt{ {\rm eigenval}\left[ (\mathbb{I}-\mathcal{C}_{\Pi}^{-1}\mathcal{C}_{p2} )(\mathbb{I}-\mathcal{C}_{\Pi}^{-1}\mathcal{C}_{p1} ) \right]} \,.
\end{align}
This means that $Q_{{\rm D}\mathcal{E}}$ does not define a positive definite quadratic form.
The expression for the probability density of indefinite quadratic forms can be found in~\cite{mathai1992quadratic}.
Here we notice that the decomposition of $Q_{{\rm D}\mathcal{E}}$ can be written as:
\begin{align}
Q_{{\rm D}\mathcal{E}} =&\, \sum_{j=1}^{2N} \lambda_j X_j^2 = \sum_{i=1}^{N} \lambda_i X_i^2 +\sum_{j=1}^{N} \lambda_j Y_j^2 \nonumber \\
=&\,  \sum_{i=1}^{N} |\lambda_i| (X_i^2-Y_i^2)
\end{align}
where both $X$ and $Y$ are independently distributed normal variables with zero mean and unit variance and we exploited the fact that the evidence ratio has two equal eigenvalues of opposite sign.
It is now possible to show, by matching the moment-generating function, that the evidence ratio for Gaussian priors is distributed as a sum of independent variance-gamma distributed variables.
Summing all the eigenvalues shows that the distribution is zero mean and in the limit where $\mathcal{C}_{p1},\mathcal{C}_{p2}\rightarrow \mathcal{C}_\Pi$ it recovers delta prior results.

We now turn to the statistics of the ratio of likelihoods at maximum posterior.
In both the delta and uninformative flat prior cases, this follows the statistics of the data independent part of the evidence ratio.
The Gaussian case instead is given by:
\begin{align}
Q_{\rm DMAP} \equiv&\,  -2\Delta\ln \mathcal{L}_p  \\
=&\,-2 \ln \mathcal{L}_{\joint}(\theta_{p\joint}) +2 \ln \mathcal{L}_{1}(\theta_{p1}) +2 \ln \mathcal{L}_{2}(\theta_{p2})\,, \nonumber 
\end{align}
that defines a quadratic form in data space that can be easily written with Eq.~(\ref{Eq:AppMaxPostGauss}).
This quadratic form is central and positive definite and, as before, it can be written as the difference of two quadratic forms.
By direct calculation it can be shown that its eigenvalues are given by:
\begin{align} \label{Eq:EigenvaluesMPRatio}
\lambda = {\rm eigenval}
\left(\begin{array}{c c}
A  & B\\
 C &D 
 \end{array} \right) \,, 
\end{align}
where
\begin{align}
A & = \mathbb{I}-\mathcal{C}_\Pi^{-1}\mathcal{C}_{p1} -\mathcal{C}_{p1}^{-1}\mathcal{C}_{p\joint} +\mathcal{C}_{\Pi}^{-1}\mathcal{C}_{p\joint} \,, \nonumber\\
B& = \mathcal{C}_{p1}^{-1}\mathcal{C}_{p\joint} -\mathcal{C}_\Pi^{-1}\mathcal{C}_{p\joint} +\mathcal{C}_\Pi^{-1}\mathcal{C}_{p1} -\mathcal{C}_\Pi^{-1}\mathcal{C}_{p1}\mathcal{C}_\Pi^{-1}\mathcal{C}_{p1} \,,
\nonumber\\
C&=  \mathcal{C}_{p2}^{-1}\mathcal{C}_{p\joint} -\mathcal{C}_\Pi^{-1}\mathcal{C}_{p\joint} +\mathcal{C}_\Pi^{-1}\mathcal{C}_{p2} -\mathcal{C}_\Pi^{-1}\mathcal{C}_{p2}\mathcal{C}_\Pi^{-1}\mathcal{C}_{p2} \,,
\nonumber\\
D&= \mathbb{I}-\mathcal{C}_\Pi^{-1}\mathcal{C}_{p2} -\mathcal{C}_{p2}^{-1}\mathcal{C}_{p\joint} +\mathcal{C}_{\Pi}^{-1}\mathcal{C}_{p\joint} \,.
\end{align}

The quadratic form defining $Q_{\rm DMAP}$ is positive definite so that its  eigenvalues are all positive and recover the two limits of uninformative flat priors and delta priors.
Eq.~(\ref{Eq:EigenvaluesMPRatio}) can be used if one wants to compute the exact distribution of $Q_{\rm DMAP}$.
On the other hand it is convenient to approximate this distribution by a chi squared distribution, as discussed in App.~\ref{App:QuadraticForms}, with
\begin{align}
\sum \lambda = N_{\rm eff}^{1}+N_{\rm eff}^{2}-N_{\rm eff}^{\joint} \,,
\end{align}
degrees of freedom since this would generally down-weight the contribution of partially constrained parameter space directions.

\section{Optimal quadratic forms} \label{App:OptimalQuadForm}
Given that there seems to be no general rule to select the matrix defining the quadratic forms in Eq.~(\ref{EqApp:QuadraticForm}), in this appendix we discuss how to choose a quadratic form that is ``optimal'' in some sense.
For this purpose it is worth noticing that a quadratic forms defined by Eq.~(\ref{EqApp:QuadraticForm}), if rescaled by a positive quantity, $\alpha$, would give the same statistical significance of results, i.e. $P(Q > Q^{\rm obs}) = P( \alpha Q > \alpha  Q^{\rm obs})$. 
This means that, for our purpose, the quadratic forms defined by $A$ and $\alpha A$ are equivalent.

As a consequence, all quadratic forms, in one dimension give the same statistical significance.
This explains why the rule of thumb difference in mean, discussed in Sec.~\ref{Sec:ParametersQuad}, when it can be applied and is representative of the full tension, works so well. In one dimension all parameter quadratic forms are equivalent and the rule of thumb is the one for which we can immediately read the statistical significance.

In multiple dimensions the same does not apply and, apart from a constant rescaling, different choices of the matrix $A$ would lead to a different statistical significance. We follow~\cite{PhysRevE.84.031124} in looking for a quadratic form that is optimal according to some criterion.
Since the quadratic form defined by Eq.~(\ref{EqApp:QuadraticForm}) is central, i.e.~$\langle X \rangle = 0$, all the cumulants of the quadratic forms pdf are given by:
\begin{align}
\kappa_m = (m-1)! \, {\rm Tr}[ (A\Sigma)^m ] \,.
\end{align}
Starting from this, one can compute all moments. 

The mean is given by $\mu_1=\kappa_1= {\rm Tr}[ (A\Sigma) ]$ and the variance by $\mu_2=\kappa_2+\kappa_1^2= 2{\rm Tr}[ (A\Sigma)^2 ]$. For all other moments we refer the reader to~\cite{mathai1992quadratic}.

We define the optimal parameter quadratic form to minimize the variance and all other moments. This can be achieved if the quadratic form minimizes all cumulants.
The trivial solution to our optimization problem is $A=0$ which is not particularly informative and can be excluded from the solution to our problem.
We can look for other solutions by demanding that the quadratic form should not have zero mean.
Since, for our purposes, all quadratic forms that are just rescaled by a constant are equivalent we can assume that they all have the same mean, without loss of generality. Thus we need to minimize:
\begin{align}
f(A) = (m-1)! \,{\rm Tr}[ (A\Sigma)^m ] +\alpha \left[ {\rm Tr}[ (A\Sigma)] -\kappa_1\right] \,,
\end{align}
over all positive matrices $A$ and for all cumulants greater than one. 
Notice that we implemented the constraint on the average as a Lagrange multiplier $\alpha$. Taking the derivative of $f$ with respect to the Lagrange multiplier would give back the finite mean constraint.
Writing the trace in terms of the $\lambda_q$ eigenvalues of $A\Sigma$ we have:
\begin{align}
f(A) = (m-1)! \, \sum_{q=1}^N \lambda_q^m +\alpha \left[ \sum_{q=1}^N \lambda_q -\kappa_1\right]  \,,
\end{align}
that has to be minimized over all positive, non-zero $\lambda_q$.
Setting $\partial f / \partial \lambda_q=0$ we can easily find that $f$ is minimized when all $\lambda_q$ are the same, so that $A_{\rm opt}\Sigma=\mathbb{I}$ which gives:
\begin{align}
A_{\rm opt} = \Sigma^{-1} \,.
\end{align}
That is, in multiple dimensions, the quadratic form that minimizes the variance and all moments is the inverse covariance one.

To have an intuition of this result let us consider a two dimensional space and two quadratic forms $Q_{\rm opt}$ and $Q_2$. 
The first one is the optimal, inverse covariance weighted, for which $\kappa_m^{\rm opt} = 2(m-1)!$. 
The second one has a direction rescaled, with respect to the inverse covariance, by a positive constant $\lambda$ so that all cumulants are given by $\kappa_m^{2} = (\lambda^m +1)(m-1)!$. We could now say that we can make the moments of the second form arbitrarily small by properly choosing $\lambda$ but this is not taking into account invariance under rescaling. We thus rescale the second quadratic form by the ratio of the two averages, in this case $2/(\lambda+1)$ so that all cumulants are given by $\kappa_m^{2} = 2^m (\lambda^m +1)/(\lambda +1)^m (m-1)!$ and we can see that the second quadratic form has cumulants that are always bigger than the first one. 

We can now ask what happens to the statistical significance of the reported results, in our simplified example. Let's suppose that we have two uncorrelated parameters and that the observed difference between them is given by $\Delta \theta = n( \sigma_1^2, \sigma_2^2)^T$ so that $Q_{\rm opt} = 2n^2$ and $Q_2=n^2(\lambda +1)$. The eigenvalues of $A\Sigma$ in the first case are just $(1,1)$, $Q_{\rm opt}$ is chi-squared distributed with two degrees of freedom. The eigenvalues of $A\Sigma$ in the second case are given by $(\lambda, 1)$ so that $Q_2$ is the sum of a Gamma distributed and a chi-squared distributed variable. Both distributions can be easily numerically integrated to show that statistical significance is the same for $\lambda \rightarrow 1/\lambda$ and that $Q_2$, for all values of positive $\lambda$, will underestimate both confirmation biases and tensions. 
This is why we picked as a criterion for defining an optimal quadratic form the minimization of the moments higher than the mean, as this is related to a lower probability of extreme events and would thus make our concordance/discordance estimator more sensitive to the presence of tensions that might be hidden by other estimators. 

\section{Gaussian approximation of MCMC posterior}\label{App:GaussianApproximationMCMC}
In this section we describe how we approximate the posterior obtained from MCMC sampling with a multivariate Gaussian.
This approximation is useful when computing some of the statistical results of this paper and can be obtained by properly accounting for all the factors that are usually neglected when performing the sampling. 
While we adopt CosmoMC~\cite{Lewis:2002ah} conventions, similar results would apply for other samplers.

The un-normalized posterior that {\rm CosmoMC} produces can be approximated by:
\begin{align} \label{Eq:GaussianApproximationMCMC}
\ln \mathcal{P} =&\, \ln \mathcal{L}(\tilde{\theta}) -\ln V_\Pi  -\frac{1}{2}(\theta-\tilde{\theta})^T \mathcal{C}^{-1}_{\tilde{\theta}}(\theta-\tilde{\theta}) \nonumber \\
&\, +\ln \frac{\Pi(\tilde{\theta})}{\Pi_{\rm max}} \,,
\end{align}
where $\tilde{\theta}$ is the parameter around which the expansion is performed, $\mathcal{C}_{\tilde{\theta}} = \langle (\theta-\tilde{\theta})(\theta-\tilde{\theta}) \rangle_{\theta}$ the covariance of the parameters samples around that point and $\mathcal{L}(\tilde{\theta})$ the likelihood at that point.
We also included a prior term that takes into account that some parameters, i.e.~some nuisance parameters, might have Gaussian priors.
There are mainly three points that we can use to define our Gaussian approximation: the parameters' mean; the maximum posterior parameters; and the parameters from the maximum posterior in the MCMC samples.

It is possible to define the best Gaussian approximation by computing the KL divergence~\cite{Kullback:1951va} between the Gaussian approximation and the full posterior for the three expansion points and select the approximation that has the smallest difference in information content with respect to the full posterior.

Having $N$ samples $\theta_i$ of the parameter posterior the KL divergence, $D_{\rm KL}$, between the (normalized) full posterior, $P_{\rm full}$, and one of the Gaussian approximations, $P_{\rm G}$, can be written as:
\begin{align} \label{Eq:KLdivergenceSamples}
D_{\rm KL}( P_{\rm full} || P_{\rm G} ) \equiv\,& \int P_{\rm full}(\theta)  \ln \left[ \frac{P_{\rm full}(\theta) }{P_{\rm G}(\theta) } \right] \, d\theta \nonumber \\
\simeq\,& \frac{1}{N} \sum_{i=1}^{N}  \ln \left[ \frac{ \mathcal{P}_{\rm full}(\theta_i) }{\mathcal{P}_{\rm G}(\theta_i)}  \right] +C
\end{align}
where the samples $\theta_i$ are drawn from $P_{\rm full}$ and $\mathcal{P}_{\rm G}$ is easily computed with Eq.~(\ref{Eq:GaussianApproximationMCMC}).
The normalization constant $C$ is the ratio between the evidence of the full posterior and the evidence of the Gaussian approximation $C=\ln ( \mathcal{E}_{\rm G} / \mathcal{E}_{\rm full})$.
Notice that, for the purpose of comparing performances of different Gaussian approximations, there is no need for an accurate estimate of the full posterior evidence.
Eq.~(\ref{Eq:KLdivergenceSamples}) is trivially generalized to weighted samples.

Given the Gaussian approximation of the MCMC posterior we can compute the evidence as:
\begin{align} \label{Eq:GaussianEvidenceMCMC}
\ln \mathcal{E} =&\, \ln \mathcal{L}(\tilde{\theta}) -\ln V_\Pi +\frac{N}{2}\ln(2\pi) +\frac{1}{2} \ln | \mathcal{C}_{\tilde{\theta}} | \nonumber \\
&\, +\ln\frac{\Pi(\tilde{\theta})}{\Pi_{\rm max}}  \,,
\end{align}
which is usually called the Laplace or saddle-point approximation and we accounted for Gaussian priors on some parameters.

In general one cannot test whether a distribution is truly Gaussian but we can perform several null tests to warn us against non-Gaussianities in parameter space.
In particular we checked:
\begin{itemize}
\item that the marginalized 1D posterior was visually well approximated by the marginalized 1D Gaussian approximation, for all constrained parameters;
\item that the best fit obtained by explicitly minimizing the data residuals, the best fit from MCMC samples and the mean were not showing relevant shifts in units of their covariance, for all the constrained parameters;
\end{itemize}
Whenever one of the Gaussian approximations fails to comply with these requirements we flag the results and express caution in interpreting them.

\section{Parameters priors}\label{App:Priors}
\begin{table}[!th]
\setlength{\tabcolsep}{12pt}
\centering
\begin{tabular}{@{}llllll@{}}
\toprule
Parameter & Prior range  \\
\colrule
$\Omega_b h^2$          & $[\, 0.005 \,,\, 0.1  \,]$ \\[1mm]
$\Omega_c h^2$          & $[\, 0.001 \,,\, 0.99 \,]$   \\[1mm]
$ 100\theta_{\rm MC}$        & $[\, 0.5 \,,\, 10 \,]$     \\[1mm]
$\tau$                   & $[\, 0.01 \,,\, 0.8 \,]$        \\[1mm]
$n_s$                     & $[\, 0.8 \,,\, 1.2 \,]$     \\[1mm]
${\rm{ln}}(10^{10} A_s)$  & $[\, 2 \,,\, 4 \,]$         \\
\botrule
\end{tabular}
\caption{
\label{Table:Priors}
Nominal flat priors on the six cosmological parameters of the $\Lambda$CDM model used for all analyses in this work.
}
\end{table}

The estimate of most of the results in the main text depends on the prior, especially in quantifying how many directions a data set constrains compared to it. In this appendix we discuss how we approximate the prior distribution.

In many cases these are informative flat priors and we approximate them with Gaussian priors of the same covariance to compute the statistics discussed in the main text while taking into account that explicit evaluations of the prior would give $\Pi(\theta)=1/V_\Pi$.  In order to make these approximations we sample the parameter space for the various flat priors listed in Table~\ref{Table:Priors} to obtain the covariance and volume.
To make our approach more efficient and transparent we do not sample Gaussian priors but rather account for their variance analytically as described below.

\begin{figure}[!th]
\centering
\includegraphics[width=\columnwidth]{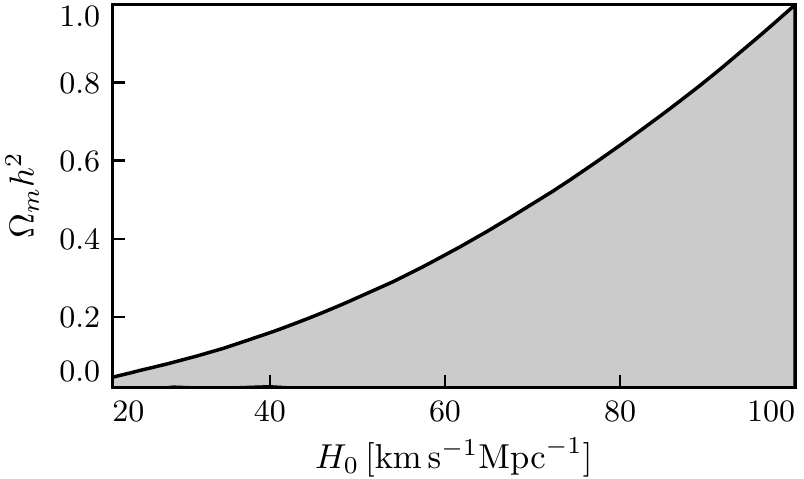}	
\caption{\label{Fig:PriorMarginal2D}
The two dimensional marginalized prior distribution of physical matter density $\Omega_m h^2$ and the Hubble constant $H_0$. 
The shaded area shows parameter choices that satisfy the constraints $0 \leq \Omega_{m} \leq 1$ and $20 \leq H_0\, [{\rm km}\,{\rm s}^{-1}{\rm Mpc}^{-1}] \leq 100$.
This projection highlights that priors on derived quantities leave the prior distribution flat but introduce a non-trivial shape to the boundaries of the prior volume.
}
\end{figure}

While approximate, this approach works very well in practice. It is faster and computationally less expensive than re-sampling the parameter posterior and is less noisy with respect to the results obtained by importance sampling the MCMC samples with a Gaussian prior.
Its robustness stems from the fact that the most important information that we need to extract from the prior is whether a parameter is constrained or not.
Other situations that fall in between are not usually relevant to the end results.

In addition to the parameters in Table~\ref{Table:Priors}, the likelihood of most experiments will add nuisance parameters describing systematic effects. 
We include them by analytically augmenting the prior covariance matrix and volume.
In case of flat priors on nuisance parameters we fill the corresponding entrance in the prior covariance with $\mathcal{C} = (\theta_{\rm max} -\theta_{\rm min})^2/12$ which corresponds to the variance of the flat distribution between $\theta_{\rm max}$ and $\theta_{\rm min}$.
Some nuisance parameters, noticeably some foreground parameters in CMB observations, have tight uncorrelated Gaussian priors. 
In this case the corresponding prior covariance entrance can be easily read from the parameter prior variance.

The prior on the base $\Lambda$CDM parameters deserves a closer look. We choose a parameter basis that has $ 100\theta_{\rm MC}$ instead of the Hubble constant but we also impose physical constraints on matter density $\Omega_{m}$ to be positive definite and smaller than unity.
Furthermore we impose a prior cut on the Hubble constant to be between $20\, {\rm km}\,{\rm s}^{-1}{\rm Mpc}^{-1}$ and $100\, {\rm km}\,{\rm s}^{-1}{\rm Mpc}^{-1}$.

These two joint boundary constraints on derived quantities make the prior volume non-trivial in shape and the prior covariance matrix non-diagonal in the base parameters. 
In Fig.~\ref{Fig:PriorMarginal2D} we show the 2D marginalized distribution of the prior to show that the two constraints on derived parameters are still locally flat in the interior, but the shape of the boundary induces a covariance between the parameters.

\begin{figure}[!th]
\centering
\includegraphics[width=\columnwidth]{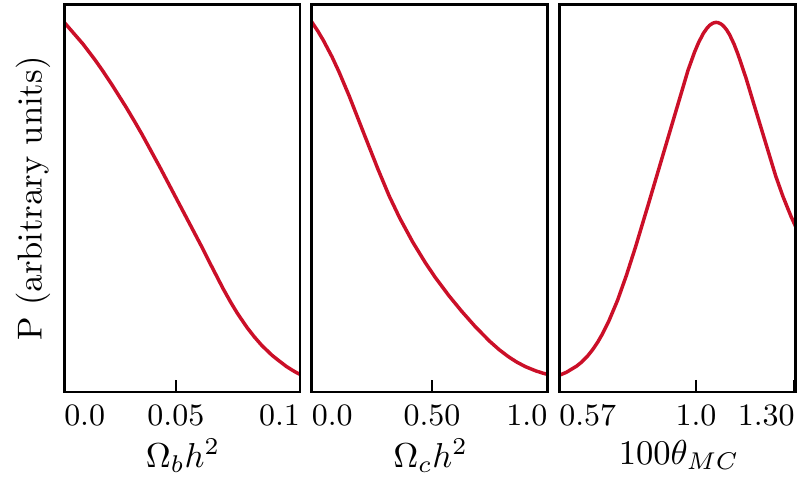}	
\caption{\label{Fig:PriorMarginal1D}
The one dimensional marginalized prior distribution on the background $\Lambda$CDM parameters.
Notice that the actual range of $ 100\theta_{\rm MC}$ is much smaller than the nominal one reported in Tab.~\ref{Table:Priors}.
}
\end{figure}
When we marginalize these flat but shaped priors to obtain the marginal distributions in 1D on the three $\Lambda$CDM background parameters, we obtain Fig.~\ref{Fig:PriorMarginal1D}.
As we can see the prior distribution for $\Omega_{b}h^2$ and $\Omega_{c}h^2$ looks to have curvature on the same scale of the prior range while the prior on $ 100\theta_{\rm MC}$ is more constraining.
This shape of the 1D prior will not influence the posterior distribution for constraining data sets as the prior is locally flat but 
does change the parameter ranges and combinations out to which the prior influences weaker data constraints.
In particular the range of $100\theta_{\rm MC}$ is modified, with respect to its face value in Table~\ref{Table:Priors}, and not taking that into account would lead to wrong degree of freedom counting, for data sets that do not constrain it.

\begin{figure}[!bt]
\centering
\includegraphics[width=\columnwidth]{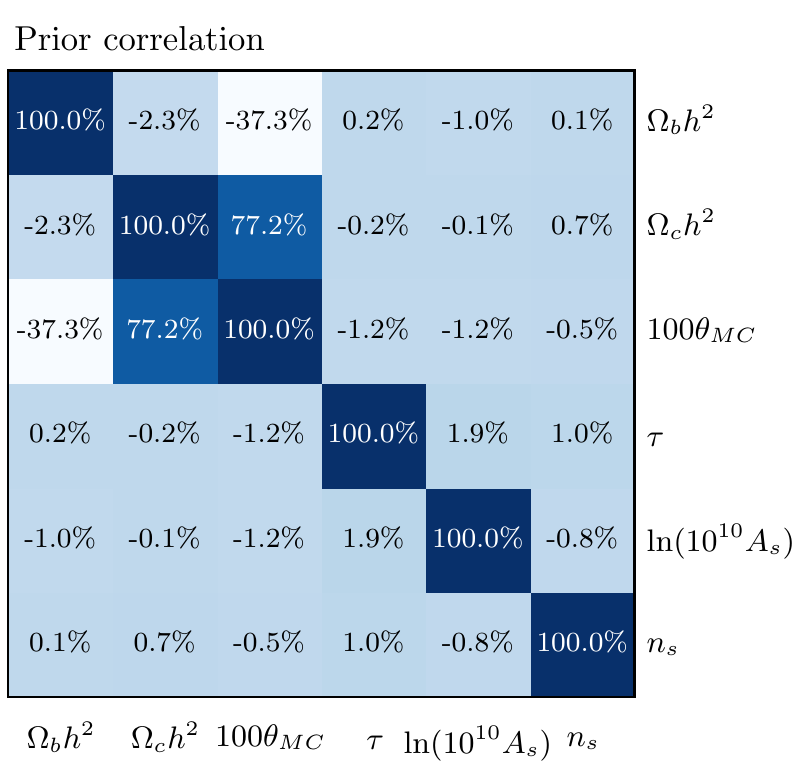}	
\caption{\label{Fig:PriorCorrelation}
The correlation between the base $\Lambda$CDM parameter priors, as obtained form the prior MCMC samples.
Notice the correlation between background parameters induced by the boundaries of the prior volume.
}
\end{figure}

Moreover, we show in Fig.~\ref{Fig:PriorCorrelation}, the prior correlation between different parameters to highlight that the prior on the background ones are also correlated because of the non-trivial shape induced by priors on derived quantities.
This correlation too is important when judging parameters shifts and counting degrees of freedom.

The remaining three  parameters $\{ \tau, n_s, \ln A_s\}$  have flat distributions and no covariance between themselves or the other parameters.
Their covariance is also well approximated by the covariance of the uniform distribution as $\mathcal{C} = (\theta_{\rm max} -\theta_{\rm min})^2/12$.
Small correlation values in Fig.~\ref{Fig:PriorCorrelation} are due to the MCMC sampling.

In summary, throughout this work we use a  Gaussian approximation to the prior for the six base parameters by using the covariance extracted from the prior MCMC samples.

\section{Tables of results}\label{App:Tables}

In this appendix we report the full results of the application of the CDEs in Sec.~\ref{Sec:Methodology} to cosmological data, in table format.
Specifically we report: 
exact 1D parameter shifts, $T_1$, and the ``rule of thumb difference in mean'', as the Gaussian approximation of $T_1$, described in Sec.~\ref{Sec:ParametersQuad} and~\ref{Sec:ResParameters}, in Table~\ref{Table:BenchmarkTension};
the likelihood at maximum posterior goodness of fit $Q_{\rm MAP}$, described in Sec.~\ref{Sec:GoF} and~\ref{Sec:ResGoF}, in Table~\ref{Table:GoFresults}; 
the difference of log-likelihoods at maximum posterior $Q_{\rm DMAP}$, described in Sec.~\ref{Sec:EvidenceRatio} and~\ref{Sec:ResEvidence}, in Table~\ref{Table:EvidenceRatioResults};
the parameter update $Q_{\rm UDM}$, described in Sec.~\ref{Sec:ParametersQuad} and~\ref{Sec:ResParameters}, in Table~\ref{Table:ParameterDifferenceResults}.

In this appendix we also report probabilities ($P$) in terms of equivalent number of standard deviations ($n_\sigma$). This should be interpreted as an effective definition corresponding to a Gaussian distribution:
\begin{align} \label{Eq:EffectiveNumberSigma}
n_{\sigma}^{\rm eff}(P) \equiv \sqrt{2}{\rm Erf}^{-1} ( 1- {\rm min}[P,1-P]) \,,
\end{align}
where ${\rm Erf}^{-1}$ is the inverse error function.  
Notice that by defining the correspondence with ${\rm min}[P,1-P]$ instead of $2\,{\rm min}[P,1-P]$ as in Eq.~(\ref{Eq:T1twotailed}) we are equating the tension and confirmation tails of the non-Gaussian CDE distribution separately to the sum of probabilities in the two tails of the Gaussian.
As an example, a tension event with probability to exceed of $P=4.55\%$ would correspond to a ``$2\sigma$" significance.
$n_{\sigma}^{\rm eff}$ should not be confused with the number of standard deviations from the mean $ (Q^{\rm obs}-\langle Q\rangle) /\sqrt{{\rm Var}(Q)}$.

\onecolumngrid

\begin{table*}[h]
\setlength{\tabcolsep}{11pt}
\centering
\begin{tabular}{@{}llllll@{}}
\toprule
 \multirow{2}{*}{Data set $D_1$ vs.~$D_2$} &  \multirow{2}{*}{parameter} &  \multirow{2}{*}{$D_1$ result} & \multirow{2}{*}{$D_2$ result} & \multicolumn{2}{c}{$P(T_1>T_1^{\rm obs})$}  \\
& & & & ``rule of thumb" & exact 1D shift \\
\toprule
\data{LRG} vs \data{WiggleZ} & $\Omega_m$ & $0.212\pm0.043$ & $0.37\pm0.11$ & 16.0\,\% (1.4 $\sigma$) & 15.0\,\% (1.4 $\sigma$) \\
\colrule
\data{SN} vs \data{BAO} & $\Omega_m$ & $0.297\pm0.034$ & $0.358\pm0.042$ & 26.0\,\% (1.1 $\sigma$) & 28.0\,\% (1.1 $\sigma$) \\
\colrule
\data{CFHTLenS} vs \data{KiDS} & $\sigma_8 \Omega_m^{0.5}$ & $0.369\pm0.071$ & $0.281\pm0.087$ & 43.0\,\% (0.8 $\sigma$) & 43.0\,\% (0.8 $\sigma$) \\
\colrule
\data{H} vs \data{HSL} & $H_0$ & $73.0\pm1.7$ & $72.3\pm2.6$ & 82.0\,\% (1.3 $\sigma$) & 85.0\,\% (1.4 $\sigma$) \\
\colrule
\data{CMB} vs \data{H0} & $H_0$ & $67.25\pm0.73$ & $73.0\pm1.5$ & {\bf 0.073}\,\% ({\bf 3.4} $\sigma$) & {\bf 0.078}\,\% ({\bf 3.4} $\sigma$) \\
\colrule
\data{CMB} vs \data{BG} & $\Omega_m$ & $0.316\pm0.01$ & $0.32\pm0.026$ & 87.0\,\% (1.5 $\sigma$) & 87.0\,\% (1.5 $\sigma$) \\
\colrule
\data{CMB} vs \data{LRG} & $\Omega_m$ & $0.316\pm0.01$ & $0.212\pm0.043$ & {\bf 2.0}\,\% ({\bf 2.3} $\sigma$) & {\bf 4.5}\,\% ({\bf 2.0} $\sigma$) \\
\colrule
\data{CMB} vs \data{GC} & $\Omega_m$ & $0.316\pm0.01$ & $0.31\pm0.075$ & 94.1\,\% (1.9 $\sigma$) & 77.0\,\% (1.2 $\sigma$) \\
\colrule
\data{CMB} vs \data{CFHTLenS} & $\sigma_8 \Omega_m^{0.5}$ & $0.4595\pm0.0071$ & $0.369\pm0.071$ & 20.0\,\% (1.3 $\sigma$) & 20.0\,\% (1.3 $\sigma$) \\
\colrule
\data{CMB} vs \data{KiDS} & $\sigma_8 \Omega_m^{0.5}$ & $0.4595\pm0.0071$ & $0.281\pm0.087$ & {\bf 4.0}\,\% ({\bf 2.1} $\sigma$) & {\bf 2.1}\,\% ({\bf 2.3} $\sigma$) \\
\colrule
\data{CMB} vs \data{WL} & $\sigma_8 \Omega_m^{0.5}$ & $0.4595\pm0.0071$ & $0.354\pm0.058$ & 7.1\,\% (1.8 $\sigma$) & 6.7\,\% (1.8 $\sigma$) \\
\colrule
\data{BG} vs \data{GC} & $\sigma_8 \Omega_m^{0.5}$ & $0.448\pm0.03$ & $0.35\pm0.1$ & 36.0\,\% (0.9 $\sigma$) & 32.0\,\% (1.0 $\sigma$) \\
\colrule
\data{BG} vs \data{GC} & $\Omega_m$ & $0.32\pm0.026$ & $0.31\pm0.075$ & 90.0\,\% (1.6 $\sigma$) & 76.0\,\% (1.2 $\sigma$) \\
\colrule
\data{BG} vs \data{WL} & $\sigma_8 \Omega_m^{0.5}$ & $0.448\pm0.03$ & $0.354\pm0.058$ & 15.0\,\% (1.4 $\sigma$) & 16.0\,\% (1.4 $\sigma$) \\
\colrule
\data{BG} vs \data{WL} & $\Omega_m$ & $0.32\pm0.026$ & $0.3\pm0.13$ & 86.0\,\% (1.5 $\sigma$) & 70.0\,\% (1.0 $\sigma$) \\
\colrule
\data{GC} vs \data{WL} & $\sigma_8 \Omega_m^{0.5}$ & $0.35\pm0.1$ & $0.354\pm0.058$ & {\bf 96.9}\,\% ({\bf 2.2} $\sigma$) & 92.3\,\% (1.8 $\sigma$) \\
\colrule
\data{GC} vs \data{WL} & $\Omega_m$ & $0.31\pm0.075$ & $0.3\pm0.13$ & 93.1\,\% (1.8 $\sigma$) & 83.0\,\% (1.4 $\sigma$) \\
\colrule
\data{CMBTT} vs \data{lowl} & $\tau$ & $0.137\pm0.035$ & $0.067\pm0.021$ & 8.6\,\% (1.7 $\sigma$) & 9.9\,\% (1.6 $\sigma$) \\
\colrule
\data{CMBEE} vs \data{lowl} & $\tau$ & $0.191\pm0.063$ & $0.067\pm0.021$ & 6.1\,\% (1.9 $\sigma$) & 8.8\,\% (1.7 $\sigma$) \\
\colrule
\data{CMBTE} vs \data{lowl} & $\tau$ & $0.094\pm0.057$ & $0.067\pm0.021$ & 65.0\,\% (0.9 $\sigma$) & 74.0\,\% (1.1 $\sigma$) \\
\colrule
\data{CMBTTTEEE} vs \data{lowl} & $\tau$ & $0.115\pm0.026$ & $0.067\pm0.021$ & 14.0\,\% (1.5 $\sigma$) & 15.0\,\% (1.4 $\sigma$) \\
\botrule
\end{tabular}
\caption{ \label{Table:BenchmarkTension}
\footnotesize
The ``rule of thumb difference in mean'' and 1D exact parameter shift estimators applied to different data sets and data sets combinations. The second column indicates the parameter that is being used in the test while the third and fourth columns report its value and error for the two data sets considered. 
The last two column indicates the probability to exceed (P.T.E.) the tests and $n_\sigma^{\rm eff}$, as computed from the results of Sec.~\ref{Sec:ParametersQuad}. 
All results that are higher than $95\%$  and lower than $5\%$ P.T.E. are highlighted as statistically significant confirmation bias and tension respectively.
This table contains mostly known results that we use as a benchmark for other concordance and discordance estimators.
}
\end{table*}

\begin{samepage}
\begin{table*}[h]
\setlength{\tabcolsep}{6pt}
\centering
\begin{tabular}{@{}llllllll@{}}
\toprule
\multirow{2}{*}{Data set} &  \multirow{2}{*}{$-2\ln \mathcal{L}_{\rm MAP}$} &  \multirow{2}{*}{$N_{\rm eff}$} &  \multirow{2}{*}{$N$} &  \multirow{2}{*}{$N_{\rm data}$} & 
\multicolumn{3}{c}{$P(Q_{\rm MAP} >Q_{\rm MAP}^{\rm obs})$} \\\,$\sigma$
&&&&& min(DoF) & best(DoF)  & max(DoF) \\
\toprule
\data{CMBTT} & 757.6 & 14.3 & 21 & 765 & 57.0\,\% (0.8\,$\sigma$) & 42.0\,\% (0.8\,$\sigma$) & 36.0\,\% (0.9\,$\sigma$) \\
\colrule
\data{CMBEE} & 739.8 & 8.1 & 13 & 762 & 71.0\,\% (1.1\,$\sigma$) & 64.0\,\% (0.9\,$\sigma$) & 59.0\,\% (0.8\,$\sigma$) \\
\colrule
\data{CMBTE} & 924.6 & 7.9 & 15 & 762 & 0.0045\,\% (4.1\,$\sigma$) & {\bf 0.0019}\,\% ({\bf 4.3}\,$\sigma$) & 0.00089\,\% (4.4\,$\sigma$) \\
\colrule
\data{CMBL} & 5.3 & 2.5 & 7 & 8 & 73.0\,\% (1.1\,$\sigma$) & 44.0\,\% (0.8\,$\sigma$) & 2.1\,\% (2.3\,$\sigma$) \\
\colrule
\data{CMBTTTEEE} & 2417.1 & 19.0 & 33 & 2289 & 3.1\,\% (2.2\,$\sigma$) & {\bf 1.6}\,\% ({\bf 2.4}\,$\sigma$) & 0.93\,\% (2.6\,$\sigma$) \\
\colrule
\data{SN} & 695.1 & 3.0 & 8 & 740 & 88.0\,\% (1.6\,$\sigma$) & 86.0\,\% (1.5\,$\sigma$) & 83.0\,\% (1.4\,$\sigma$) \\
\colrule
\data{BAO} & 5.4 & 3.1 & 6 & 11 & 90.9\,\% (1.7\,$\sigma$) & 70.0\,\% (1.0\,$\sigma$) & 37.0\,\% (0.9\,$\sigma$) \\
\colrule
\data{LRG} & 4.1 & 2.5 & 6 & 14 & 99.49\,\% (2.8\,$\sigma$) & {\bf 97.5}\,\% ({\bf 2.2}\,$\sigma$) & 85.0\,\% (1.4\,$\sigma$) \\
\colrule
\data{WiggleZ} & 189.5 & 1.9 & 6 & 196 & 62.0\,\% (0.9\,$\sigma$) & 58.0\,\% (0.8\,$\sigma$) & 50.0\,\% (0.7\,$\sigma$) \\
\colrule
\data{CFHTLenS} & 86.8 & 1.8 & 7 & 56 & 0.52\,\% (2.8\,$\sigma$) & {\bf 0.32}\,\% ({\bf 2.9}\,$\sigma$) & 0.07\,\% (3.4\,$\sigma$) \\
\colrule
\data{KiDS} & 58.4 & 1.8 & 7 & 30 & 0.14\,\% (3.2\,$\sigma$) & {\bf 0.07}\,\% ({\bf 3.4}\,$\sigma$) & 0.0064\,\% (4.0\,$\sigma$) \\
\botrule
\data{CMB} & 2432.2 & 18.9 & 33 & 2297 & 2.5\,\% (2.2\,$\sigma$) & {\bf 1.2}\,\% ({\bf 2.5}\,$\sigma$) & 0.71\,\% (2.7\,$\sigma$) \\
\colrule
\data{BG} & 702.2 & 5.0 & 8 & 751 & 90.0\,\% (1.6\,$\sigma$) & 87.0\,\% (1.5\,$\sigma$) & 86.0\,\% (1.5\,$\sigma$) \\
\colrule
\data{GC} & 204.7 & 2.5 & 6 & 210 & 59.0\,\% (0.8\,$\sigma$) & 54.0\,\% (0.7\,$\sigma$) & 47.0\,\% (0.7\,$\sigma$) \\
\colrule
\data{WL} & 146.5 & 2.7 & 8 & 86 & 0.0052\,\% (4.0\,$\sigma$) & {\bf 0.0024}\,\% ({\bf 4.2}\,$\sigma$) & 0.00044\,\% (4.6\,$\sigma$) \\
\colrule
\data{ALL} & 3516.2 & 22.8 & 37 & 3345 & 1.9\,\% (2.3\,$\sigma$) & {\bf 0.96}\,\% ({\bf 2.6}\,$\sigma$) & 0.59\,\% (2.8\,$\sigma$) \\
\botrule
\end{tabular}
\caption{ \label{Table:GoFresults}
\footnotesize
The likelihood at maximum posterior (MAP) goodness of fit estimator applied to different data sets and  combinations. 
The second column reports the data likelihood at maximum posterior; the third the number of effective parameters  $N_{\rm eff}$, as estimated using 
Eq.~(\ref{Eq:Neff}); the fourth the number of nominal parameters $N$, the fifth the number of data points $N_{\rm data}=d$ and the seventh the
P.T.E. for $Q_{\rm MAP}^{\rm obs}$ assuming our best estimate of the  degrees of freedom (DoF) $N_{\rm data}-N_{\rm eff}$.  Values  higher than  $95\%$  or lower than $5\%$ P.T.E. are highlighted  as statistically significant confirmation bias and tension respectively.   The remaining columns list the P.T.E.s assuming the minimal DoF $N_{\rm data}-N$ and the maximal DoF $N_{\rm data}$ which place conservative bounds on tension and confirmation respectively.
}
\end{table*}
\begin{table*}[h]
\setlength{\tabcolsep}{10pt}
\centering
\begin{tabular}{@{}llllllll@{}}
\toprule
Data set & $N^{\rm eff}_{1}$ & $N^{\rm eff}_{2}$ & $N^{\rm eff}_{\joint}$ & $\Delta N^{\rm eff}$ & $\log_{10}\mathrm{C}$ & $\langle\log_{10}\mathrm{C}\rangle_{\joint}$ & $P(Q_{\rm DMAP}>Q_{\rm DMAP}^{\rm obs})$ \\
\toprule
\data{LRG} vs \data{WiggleZ} & 2.5 & 1.9 & 2.5 & 1.8 & 1.0 & 3.0 & {\bf 0.31}\,\% ({\bf 3.0}\,$\sigma$) \\
\colrule
\data{SN} vs \data{BAO} & 3.0 & 3.1 & 5.0 & 1.1 & 2.4 & 2.6 & 20.0\,\% (1.3\,$\sigma$) \\
\colrule
\data{CFHTLenS} vs \data{KiDS} & 1.8 & 1.8 & 2.7 & 0.9 & 2.4 & 2.4 & 25.0\,\% (1.2\,$\sigma$) \\
\colrule
\data{CMBTT} vs \data{CMBEE} & 14.3 & 8.1 & 16.9 & 5.6 & 10.2 & 10.6 & 25.0\,\% (1.2\,$\sigma$) \\
\colrule
\data{CMBTT} vs \data{CMBL} & 14.3 & 2.5 & 14.3 & 2.5 & 3.3 & 4.8 & {\bf 1.8}\,\% ({\bf 2.4}\,$\sigma$) \\
\colrule
\data{CMBEE} vs \data{CMBL} & 8.1 & 2.5 & 8.4 & 2.3 & 2.6 & 4.1 & {\bf 1.4}\,\% ({\bf 2.5}\,$\sigma$) \\
\colrule
\data{CMBTE} vs \data{CMBL} & 7.9 & 2.5 & 8.0 & 2.4 & 3.9 & 4.2 & 18.0\,\% (1.3\,$\sigma$) \\
\colrule
\data{CMBTTTEEE} vs \data{CMBL} & 19.0 & 2.5 & 18.9 & 2.6 & 3.2 & 4.8 & {\bf 1.3}\,\% ({\bf 2.5}\,$\sigma$) \\
\botrule
\data{CMB} vs \data{BG} & 18.9 & 5.0 & 21.0 & 3.0 & 4.2 & 3.8 & 75.0\,\% (1.2\,$\sigma$) \\
\colrule
\data{CMB} vs \data{GC} & 18.9 & 2.5 & 18.9 & 2.6 & 2.2 & 3.5 & {\bf 2.3}\,\% ({\bf 2.3}\,$\sigma$) \\
\colrule
\data{CMB} vs \data{WL} & 18.9 & 2.7 & 20.8 & 0.8 & 0.3 & 2.3 & {\bf 0.1}\,\% ({\bf 3.3}\,$\sigma$) \\
\colrule
\data{CMB} vs \data{H0} & 18.9 & 1.4 & 19.0 & 1.3 & -0.6 & 2.5 & {\bf 0.088}\,\% ({\bf 3.3}\,$\sigma$) \\
\colrule
\data{BG} vs \data{GC} & 5.0 & 2.5 & 5.5 & 2.1 & 1.8 & 3.1 & {\bf 2.3}\,\% ({\bf 2.3}\,$\sigma$) \\
\colrule
\data{BG} vs \data{WL} & 5.0 & 2.7 & 6.9 & 0.9 & 2.0 & 2.3 & 12.0\,\% (1.6\,$\sigma$) \\
\colrule
\data{GC} vs \data{WL} & 2.5 & 2.7 & 4.3 & 0.9 & 1.6 & 2.9 & {\bf 0.74}\,\% ({\bf 2.7}\,$\sigma$) \\
\colrule
\data{GC} vs \data{H0} & 2.5 & 1.4 & 3.2 & 0.7 & 1.3 & 1.6 & 9.7\,\% (1.7\,$\sigma$) \\
\colrule
\data{WL} vs \data{H0} & 2.7 & 1.4 & 3.6 & 0.4 & 1.9 & 1.9 & 22.0\,\% (1.2\,$\sigma$) \\
\botrule
\end{tabular}
\caption{ \label{Table:EvidenceRatioResults}
\footnotesize
Evidence ratio type estimators applied to different data sets combinations. 
The first three columns report the number of effective parameters of the first, second and joint data sets.
The fourth column reports the number of effective parameters that both data sets constrain.
The fifth column reports the observed value of the evidence ratio and the sixth one its expected value when averaged over data realizations of $D_1 \cup D_2$.
The last column reports the significance of the observed value of the ratio of likelihoods at maximum posterior (DMAP), as estimated using the results of Sec.~\ref{Sec:EvidenceRatio}.
All results that are higher than  $95\%$  and lower than $5\%$ P.T.E. are highlighted as statistically significant confirmation bias and tension respectively.
}
\end{table*}
\end{samepage}

\begin{table*}[h]
\setlength{\tabcolsep}{12pt}
\centering
\begin{tabular}{@{}llll@{}}
\toprule
Data set & $Q_{\rm UDM}$ & $N_{\rm KL}$ & $P(Q_{\rm UDM}>Q_{\rm UDM}^{\rm obs})$ \\
\toprule
\data{LRG} vs \data{WiggleZ} & 5.5 & 1 & {\bf 1.9}\,\% ({\bf 2.3}\,$\sigma$) \\
 \colrule
\data{BAO} vs \data{SN} & 1.0 & 1 & 33.0\,\% (1.0\,$\sigma$) \\
 \colrule
\data{CFHTLenS} vs \data{KiDS} & 0.1 & 1 & 75.0\,\% (1.2\,$\sigma$) \\
 \colrule
\data{H} vs \data{HSL} & 0.3 & 1 & 62.0\,\% (0.9\,$\sigma$) \\
 \colrule
\data{CMBTT} vs \data{CMBL} & 7.0 & 1& {\bf 0.82}\,\% ({\bf 2.6}\,$\sigma$) \\
 \colrule
\data{CMBEE} vs \data{CMBL} & 6.6 & 1 & {\bf 1.0}\,\% ({\bf 2.6}\,$\sigma$) \\
 \colrule
\data{CMBTE} vs \data{CMBL} & 0.3 & 1 & 59.0\,\% (0.8\,$\sigma$) \\
 \colrule
\data{CMBTTTEEE} vs \data{CMBL} & 7.3 & 1 & {\bf 0.68}\,\% ({\bf 2.7}\,$\sigma$) \\
 \colrule
\data{lowl} vs  \data{CMBTT} & 3.9 & 1 & {\bf 4.9}\,\% ({\bf 2.0}\,$\sigma$) \\
 \colrule
\data{lowl} vs \data{CMBEE} & 8.5 & 2 & {\bf 1.4}\,\% ({\bf 2.4}\,$\sigma$) \\
 \colrule
\data{lowl} vs \data{CMBTE} & 3.2 & 2 & 20.0\,\% (1.3\,$\sigma$) \\
 \colrule
\data{lowl} vs \data{CMBTTTEEE} & 3.1 & 1 & 7.6\,\% (1.8\,$\sigma$) \\
 \colrule
\data{lowl} + \data{CMBTT} vs \data{CMBL} & 1.5 & 1 & 22.0\,\% (1.2\,$\sigma$) \\
 \colrule
\data{lowl} + \data{CMBEE} vs \data{CMBL} & 1.1 & 2 & 59.0\,\% (0.8\,$\sigma$) \\
 \colrule
\data{lowl} + \data{CMBTE} vs \data{CMBL} & 0.1 & 1 & 77.0\,\% (1.2\,$\sigma$) \\
 \colrule
\data{lowl} + \data{CMBTTTEEE} vs \data{CMBL} & 2.0 & 1 & 16.0\,\% (1.4\,$\sigma$) \\
 \colrule
\data{lowl} vs  \data{CMBTT} + \data{CMBL} & 0.0 & 1 & 88.0\,\% (1.6\,$\sigma$) \\
 \colrule
\data{lowl} vs \data{CMBEE} + \data{CMBL} & 2.5 & 2 & 29.0\,\% (1.1\,$\sigma$) \\
 \colrule
\data{lowl} vs \data{CMBTE} + \data{CMBL} & 3.1 & 2 & 22.0\,\% (1.2\,$\sigma$) \\
 \botrule
\data{CMB} vs \data{BG} & 0.4 & 1 & 52.0\,\% (0.7\,$\sigma$) \\
 \colrule
\data{CMB} vs \data{GC} & 0.0 & 0 & $-$ \\
 \colrule
\data{CMB} vs \data{WL} & 5.8 & 1 & {\bf 1.6}\,\% ({\bf 2.4}\,$\sigma$) \\
 \colrule
\data{CMB} vs \data{H0} & 11.1 & 1 & {\bf 0.087}\,\% ({\bf 3.3}\,$\sigma$) \\
 \colrule
\data{lowl} + \data{CMB} vs \data{BG} & 0.4 & 1 & 55.0\,\% (0.8\,$\sigma$) \\
 \colrule
\data{lowl} + \data{CMB} vs \data{GC} & 0.0 & 0 & $-$  \\
 \colrule
\data{lowl} + \data{CMB} vs \data{WL} & 5.9 & 1 & {\bf 1.5}\,\% ({\bf 2.4}\,$\sigma$) \\
 \colrule
\data{lowl} + \data{CMB} vs \data{H0} & 10.7 & 1 & {\bf 0.11}\,\% ({\bf 3.3}\,$\sigma$) \\
 \colrule
\data{BG} vs \data{GC} & 0.0 & 0 & $-$ \\
 \colrule
\data{BG} vs \data{WL} & 0.7 & 1 & 42.0\,\% (0.8\,$\sigma$) \\
 \colrule
\data{BG} vs \data{H0} & 0.4 & 1 & 55.0\,\% (0.8\,$\sigma$) \\
 \colrule
\data{GC} vs \data{WL} & 0.0 & 0 & $-$ \\
 \colrule
\data{GC} vs \data{H0} & 0.6 & 2 & 72.0\,\% (1.1\,$\sigma$) \\
 \colrule
\data{WL} vs \data{H0} & 0.2 & 2 & 90.7\,\% (1.7\,$\sigma$) \\
\botrule
\end{tabular}
\caption{ 
\label{Table:ParameterDifferenceResults}
\footnotesize
The update difference in mean estimator, $Q_{\rm UDM}$, applied to different data sets combinations.
The first column reports the observed value, as computed from Eq.~(\ref{Eq:UDMQuadraticForm}).
The second column is the number of effective KL parameters retained $N_{\rm KL} = \langle Q_{\rm UDM} \rangle_D$ for which the second data set significantly improves constraints over the first one.  
The third column reports the significance of the observed value of the update difference in mean, as estimated using the results of Sec.~\ref{Sec:ParametersQuad}.
All results that are higher than $95\%$  and lower than $5\%$ P.T.E. are highlighted as statistically significant confirmation bias and tension respectively.
When $N_{\rm KL} =0$, $Q_{\rm UDM}=0$ and we do not report statistical significance.
}
\end{table*}

\clearpage
\bibliography{biblio}

\end{document}